\documentclass[prl,twocolumn,showpacs,preprintnumbers,amsmath,amssymb,superscriptaddress]{revtex4-2}

\usepackage{graphicx}
\usepackage{dcolumn}
\usepackage{bm}
\usepackage{color}
\usepackage{bbm}
\usepackage[colorlinks=true,citecolor=blue,linkcolor=blue,urlcolor=blue]{hyperref}
\begin{document}
\title{Fast generation of high-fidelity mechanical non-Gaussian states via additional amplifier and photon subtraction}
\author{Dong-Long Hu}
\affiliation{School of Physics, Sun Yat-sen University, Guangzhou 510275, China }
\author{Jia-Jin Zou}
\affiliation{School of Physics, Sun Yat-sen University, Guangzhou 510275, China }
\author{Feng-Xiao Sun}
\affiliation{State Key Laboratory for Mesoscopic Physics, School of Physics, Frontiers Science Center for Nano-Optoelectronics, and Collaborative Innovation Center of Quantum Matter, Peking University, Beijing 100871, China}
\author{Jie-Qiao Liao}
\affiliation{Key Laboratory of Low-Dimensional Quantum Structures and Quantum Control of Ministry of Education,
	Key Laboratory for Matter Microstructure and Function of Hunan Province, Department of Physics and
	Synergetic Innovation Center for Quantum Effects and Applications, Hunan Normal University, Changsha 410081, China}
\author{Qiongyi He}
\affiliation{State Key Laboratory for Mesoscopic Physics, School of Physics, Frontiers Science Center for Nano-Optoelectronics, and Collaborative Innovation Center of Quantum Matter, Peking University, Beijing 100871, China}
\author{Ze-Liang Xiang}
\email{xiangzliang@mail.sysu.edu.cn}
\affiliation{School of Physics, Sun Yat-sen University, Guangzhou 510275, China }
\date{\today}

\begin{abstract}
	Non-Gaussian states (NGSs) with higher-order correlation properties have wide-range applications in quantum information processing. However, the generation of such states with high quality still faces practical challenges. Here, we propose a protocol to faithfully generate two types of mechanical NGSs, i.e., Schr\"{o}dinger cat states and Fock states, in open optomechanical systems, even when the cooperativity is smaller than one ($g^2/\kappa\gamma<1$). In contrast to the usual scheme, a short squeezed field is pumped to rapidly entangle with a mechanical resonator via a beam-splitter-like optomechanical interaction, effectively reducing the mechanical decoherence. Furthermore, by performing an additional amplifier and a following multi photon subtraction on the entangled optical field, one can selectively obtain the high-fidelity mechanical cat and Fock states. This protocol is robust to various imperfections, allowing it to be implemented with state-of-the-art experimental systems with close to unit fidelity. Moreover, it can be extended to generate a four-component cat state and provide possibilities for future quantum applications of NGSs.
\end{abstract}
\maketitle
% ---------------------------------------------------------------------------
% ---------------------------------------------------------------------------

%\emph{Introduction.}---
Non-Gaussian states with Wigner negativity are of paramount importance for their advantages in quantum information processing, which cannot be simulated by classical resources~\cite{Kitagawa2006,Menicucci2006,Ourjoumtsev2007b,Takahashi2010,Mari2012}. Such states, including Schr\"{o}dinger cat states (CSs) and Fock states, are widely applied in quantum error correction~\cite{Ofek2016,Ma2020}, quantum metrology~\cite{Zurek2001,Joo2011}, and quantum sensing~\cite{Manuel2019,Wolf2019}. By now, one can prepare NGSs by using four-wave mixing with Kerr nonlinearities in superconducting circuits~\cite{Leghtas2015,Ofek2016,Ma2020} or by performing non-Gaussian operations, such as photon subtraction or addition, on given Gaussian states in optical systems~\cite{Wenger2004,Zavatta2004,Neergaard2006,Ourjoumtsev2007,Wakui2007,Parigi2007,Ra2020}. In addition, NGSs can be remotely generated between distant sites through the shared entanglement in optical and microwave systems offering intrinsic security and efficiency~\cite{Lo2000,Paris2003,LeJeannic2018,Pogorzalek2019}. 

Among different systems, the well-studied cavity optomechanical system is a promising candidate for studying NGSs~\cite{Aspelmeyer2014,Kurizki2015}. This system, driven by the radiation pressure, provides an important means to manipulate and detect mechanical motion in the quantum regime using light, hence it has received substantial attention and intensive investigations. So far, many fundamental quantum phenomena of mechanical motion have been experimentally observed in this system, such as optical-mechanical~\cite{Riedinger2016,fiaschi2021,Marinkovic2018} and mechanical-mechanical entanglement~\cite{Riedinger2018,Wollack2022}, and mechanical squeezing~\cite{Rashid2016,Wollman2015}. Fortunately, these results are incredibly favorable for preparing the mechanical NGSs. Based on optomechanical systems, various schemes have been proposed for the controllable generation of mechanical NGSs with decent quality~\cite{Lv2013,Liao2016,khosla2018,hauer2022,gely2021}. However, the intrinsic mechanical dissipation and the challenging realization of strong single-photon optomechanical coupling still hinder the practical implementation of such schemes.

% ---------------------------------------------------------------------------
\begin{figure}[tbp]
	\centering
	\includegraphics[width=1\linewidth]{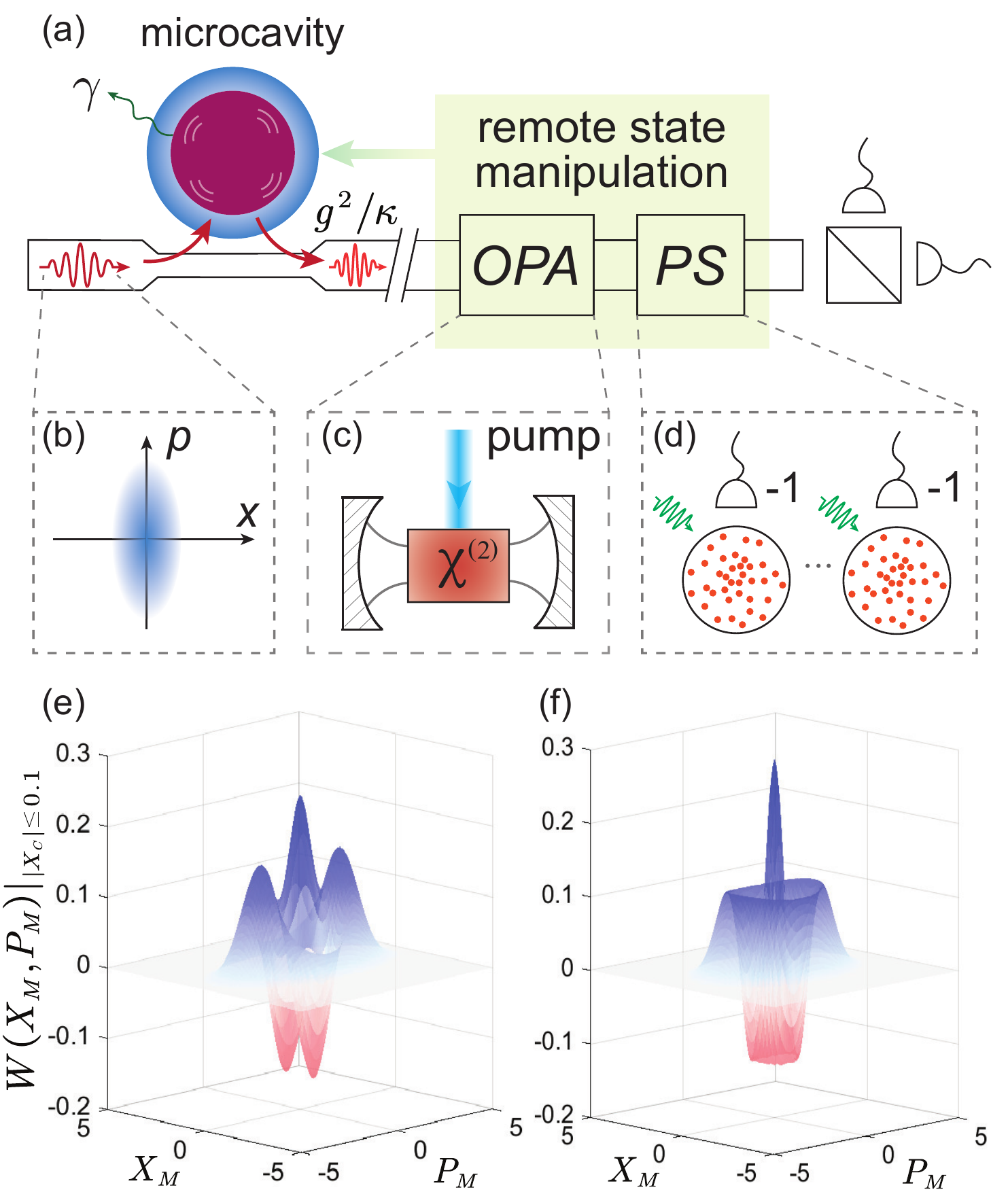}
	\caption{(a) Sketch of the fast generation of mechanical NGSs, where a short squeezed pulse interacts with the mechanical resonator and propagates to a distant site. By performing an EPS and a further projective measurement on the entangled optical mode, the desired mechanical NGSs can be faithfully produced. (b) Schematic of a squeezed vacuum. (c) Schematic of an optical parametric amplifier (OPA) and (d) schematic of a controlled multi photon subtraction with cascaded Rydberg atom ensembles. (e), (f) Resulting mechanical NGSs, a cat state and a Fock state (fidelities $\sim0.99$), with different gains of a two-photon EPS, respectively.}
	\label{fig1}
\end{figure}
% ---------------------------------------------------------------------------

In this Letter, we present a protocol for faithfully preparing near-perfect CSs and Fock states of mechanical motion in an open optomechanical system, as shown in Fig.~\ref{fig1}(a). Unlike previous works~\cite{Sun2021,Takase2021}, here a short squeezed vacuum, as shown in Fig.~\ref{fig1}(b), is pulsed to rapidly entangle with a mechanical resonator via a beam-splitter-like optomechanical interaction. This effectively reduces the mechanical decoherence and can be achieved even when cooperativity is smaller than one ($g^2/\kappa\gamma<1$). Importantly, an engineered photon subtraction (EPS) is  then performed on the entangled optical mode, which consists of a phase-sensitive amplifier followed by a multi photon subtraction, as shown in Figs.~\ref{fig1}(c) and \ref{fig1}(d). There always exists appropriate gains of the amplifier to ensure the Wigner negativity of the mechanical mode after photon subtraction and projective measurement. Interestingly, specific gains of the amplifier enable the selective preparation of high-fidelity mechanical CSs and Fock states, as shown in Figs.~\ref{fig1}(e) and \ref{fig1}(f). In addition, the significant experimental progress of deterministic photon subtraction utilizing the Rydberg-blockade effect~\cite{Honer2011,Tresp2016,Stiesdal2021} ensures that our protocol can efficiently generate large-size NGSs. Furthermore, this protocol can be extended to prepare a four-component cat state with high fidelity and provides a possibility to explore advanced NGSs. All the results here are obtained using the accessible parameters from state-of-the-art experiments.
% ---------------------------------------------------------------------------
% ---------------------------------------------------------------------------

\emph{Cavity optomechanical system.} We consider a typical cavity optomechanical system, consisting of a mechanical mode of a resonator coupled to an optical mode of a cavity via the radiation pressure~\cite{Aspelmeyer2014}. The mechanical mode is initially prepared in the ground state at a low environment temperature~\cite{Mirhosseini2020}. The dynamics of the system can be described by $H=\omega_mm^\dag m+\omega_c c^\dag c+ g_0c^\dag c(m+m^\dag)$, where $m~(c)$ is the annihilation operator of the mechanical (optical) mode with the resonate frequency $\omega_{m(c)}$, $g_0$ denotes the single-photon optomechanical coupling rate, and $\hbar = 1$. Under the red-detuned drive on the optical mode, the effective Hamiltonian reduces to (see the Supplemental Material~\cite{SM})
\begin{equation}
	H_{\rm eff}=\omega_mm^\dag m+\Delta c^\dag c-i g(m c^\dag-m^\dag c),
\end{equation}
where $\Delta=\omega_c-\omega_p$ is the frequency detuning with $\omega_p$ the frequency of the driving field, and $g=g_0|\beta|$ is the linearized optomechanical coupling rate with $|\beta|$ the light amplitude of the cavity mode. This interaction describes a beam-splitter-type scattering between the optical photons and the mechanical phonons. Unlike previous schemes that require a strong optomechanical coupling, i.e., $C_{\rm om}=g^2/\kappa\gamma>1$, where $\gamma$ represents the mechanical dissipation rate and $\kappa$ denotes the decay rate of the optical cavity, our protocol is also valid in the weak-coupling regime. To show this, here we consider a low optomechanical cooperativity $C_{\rm om}=0.8$, with the parameters $g/2\pi=3$, $\kappa/2\pi=7$, and $\gamma/2\pi=1.6$ MHz, which can be easily achieved in current experiments~\cite{Patel2021,Enzian2021,Verhagen2012,Shandilya2021}.

Because the optical cavity decay rate $\kappa\gg g$, the cavity field is adiabatically eliminated, resulting in a direct interaction between the mechanical mode and the propagating squeezed vacuum field in a waveguide, as described by the input-output relation $c_{\rm out}=c_{\rm in} +\sqrt{2\kappa}c$~\cite{Hofer2011,Vanner2011}. This interaction gives rise to an effective decay rate of the mechanical mode $G={g^2}/{\kappa}+\gamma$, simultaneously, to the propagating pulse (${g^2}/{\kappa}$) and the environment ($\gamma$), as shown in Fig.~\ref{fig1} (a). Therefore, the quantum coherence feature of the mechanical mode can be protected by pumping the squeezed vacuum with a short duration
\begin{equation}
	\tau\lesssim\frac{1}{2G}~~~~(R=e^{-2G\tau}>e^{-1}),
\end{equation}
where $R$ is the reflectivity of the effective beam splitter that plays the same role of the duration $\tau$ (see Supplemental Material~\cite{SM}). In this fast entanglement preparation regime, the remote generation of mechanical NGSs is insensitive to the decoherence of the mechanical mode, which we will show in the following.

% ---------------------------------------------------------------------------
\begin{figure*}[tbp]	
	\centering
	\includegraphics[width=0.97\linewidth]{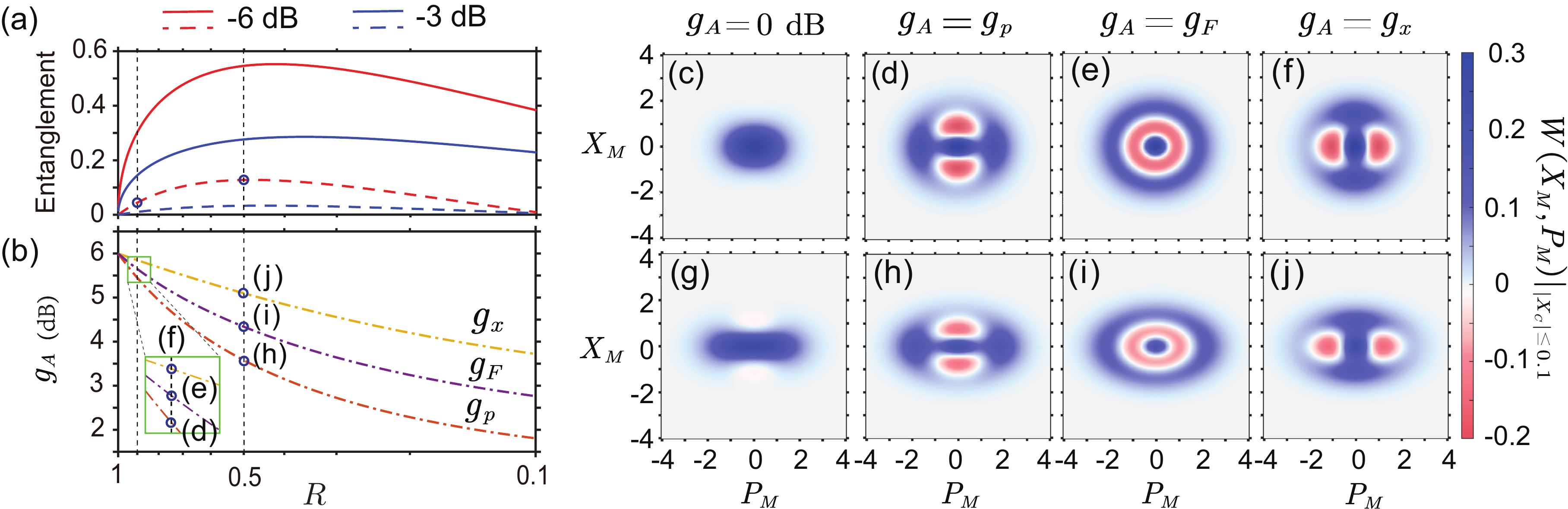}
	\caption{(a) The logarithmic negativity $E_N$ (solid line) and EPR steering $\mathcal{G}^{M\rightarrow C}$ (dashed line) vs the reflectivity $R=e^{-2G\tau}$. The red (blue) line corresponds to -6 dB (-3 dB) of the input squeezing. (b) The required gains $g_A$ of the amplifier for preparing a high-fidelity $P_M$-direction cat state ($g_p$), Fock state ($g_F$) and $X_M$-direction cat state ($g_x$) versus the reflectivity $R=e^{-2G\tau}$. (c)-(j) The resulting states by choosing the gains of the amplifier from $g_A=0$ to $g_A=g_x$ with $R=0.9$ and $R=0.5$, respectively. Here, a two-photon subtraction is adopted, and the effective mechanical decay rate $G=g^2/\kappa+\gamma=2.9$ MHz with $C_{\rm OM}=0.8$.}
	\label{fig2}
\end{figure*}
% ---------------------------------------------------------------------------
% ---------------------------------------------------------------------------

\emph{Squeezing-induced entanglement and EPR steering.} Along the squeezed vacuum propagates through and interacts with the mechanical resonator, the excepted non-local correlations between the cavity output mode ($C$) and the mechanical ($M$) mode can deterministically produce. We assume the quadrature $X_{C_{\rm in}}$ of the input pulse is squeezed with a strength $S_{\rm in}$, where the quadratures are $X_a=(a+a^\dag)/{\sqrt{2}},P_a=(a-a^\dag)/{i\sqrt{2}}$, and $a$ is a generic bosonic annihilation operator. The entanglement can be qualified by the logarithmic negativity $E_N$~\cite{Adesso2004} and the Einstein-Podolsky-Rosen (EPR) steering from the mechanical mode to the cavity output mode, $\mathcal{G}^{M\rightarrow C}$ ~\cite{Kogias2015} (see the Supplemental Material~\cite{SM}). The EPR steering is a quantum phenomenon that one party can remotely influence the wave function of the other distant party by performing suitable measurements~\cite{Uola2020}. Note that the EPR steering is a necessary resource for the remote generation of Wigner negativity in the mechanical mode via the photon subtraction~\cite{Walschaers2020,Liu2022}.  

In Fig.~\ref{fig2}(a), we illustrate the quantum correlations as functions of the reflectivity $R$ with different squeezing levels, where the photon-phonon entanglement can effectively produce with low cooperativity ($C_{\rm om}=0.8$). For a good initialization of the mechanical mode with negligible thermal occupation, arbitrary $S_{\rm in}\neq0$ dB ($x\rightarrow10\log_{10}x$ dB) is feasible for preparing effective entanglement. For example, we choose $S_{\rm in}=-$ 6 dB in the following discussions for generating high-fidelity mechanical NGSs. In experiments, the squeezing strength $S_{\rm in}\approx-15$ dB for optical modes has been achieved~\cite{Eberle2010}.
% ---------------------------------------------------------------------------
% ---------------------------------------------------------------------------

\emph{Remote state generation and manipulation.} After performing an EPS and a further projective measurement on the entangled optical mode, as shown in Fig.~\ref{fig1}(a), the mechanical mode will immediately collapse to an NGS, $\rho_M$. For investigating the precise relations between the resulting state $\rho_M$ and the EPS, we describe the $n$-photon EPS as a combined operator, $\textbf{E}(g_A,n):=c^n\mathcal{U}(g_A)$. Here the phase of the amplifier $\mathcal{U}$ matches the input squeezing, which offers a direct amplification for $X_C$ with gain $g_A$, and $n$ is the subtracted photon number from the optical mode. With the outcome $X_C=0$, the relation between $\rho_M$ and the entangled state $\rho_{out}$ is $\rho_M=\langle \textbf{E}(g_A,n)\rho_{out}\textbf{E}^{\dag}(g_A,n)\rangle_{X_C=0}.$ Ideally with $\gamma=0$, we can analytically derive the wave function $\psi(X_M)$ of $\rho_M$ in the representation of $X_M$, as
\begin{align}
	\psi(X_M)&\propto\phi_{n,\xi}(X_M)\exp\left(-\frac{X_M^2}{2\sigma_{11}^{-1}}\right),
	\label{eq3}\\
	\phi_{n,\xi}(X_M)&\propto\sigma_{13}^n\Sigma_{k=0}^{[n/2]}\frac{(-1)^kn!2^{n-2k}}{k!(n-2k)!}\left(\frac{X_M}{\sqrt{2\sigma_{11}^{-1}}}\right)^{n-2k}{\xi}^{k}.
	\label{eq4}
\end{align}
Here the non-Gaussian features of $\psi(X_M)$ are exhibited by $\phi_{n,\xi}(X_M)$, which relates to the $n$-th order Hermite polynomial $\phi_{n,\xi}(X_M)\propto H_n(X_M/\sqrt{2\xi\sigma_{11}^{-1}})$ when $\xi\neq0$. Surprisingly, we find that $\xi$ intimately relates to the non-Gaussianity of $\rho_M$ and can be remotely controlled by the applied amplifier simultaneously, i.e., $\xi=(\sigma_{33}-g_A)/(\sigma_{13}^2\sigma_{11}^{-1})$, where $\sigma=1/2V^{-1}$, $V$ is the covariance matrix of the entangled state $\rho_{\rm out}$, and $\sigma_{ij}$ is the matrix element of $\sigma$ (see Supplemental Material~\cite{SM}). 

From Eqs.~\eqref{eq3} and \eqref{eq4}, we find the following: (i) when $\xi=0$ or $\xi=1$, $\psi(X_M)$ corresponds to a near-perfect squeezed CS $\sim\hat{S}\mathcal{N}_{n}(|\alpha \rangle+(-1)^n|-\alpha \rangle)$, with a coherent amplitude $\alpha=\sqrt{n}e^{i\frac{\pi}{2}\xi}$ squeezed by $s=\sigma_{11}^{-e^{-i\pi\xi}}/2$ and a high-fidelity $\mathcal{F}\approx1-0.03/n$ (i.e., the overlap between actual and target states, defined as $\mathcal{F}=\langle\psi_t|\rho_M|\psi_t\rangle$)~\cite{Ourjoumtsev2007}; (ii) when $\xi=0.5$, a mechanical Fock state $\hat{S}|n\rangle$ squeezed by $s=\sigma_{11}^{-1}$ can be strictly generated. Here $\hat{S}$ is a formalistic squeezing operator, $\mathcal{N}_{n}$ is the normalization coefficient, and $\sigma_{11}\approx R+TS_{\rm in}^{-1}$ indicates the mechanical squeezing effect coherently transferred from the optical squeezing, which increases with the duration $\tau$~\cite{SM}. This precise mapping allows the faithful preparation of two typical mechanical NGSs, i.e., the CS and Fock state, by simply adjusting the gain $g_A$ of the amplifier. The required gain can be inversely solved by
\begin{equation}
	g_A=\sigma_{33}-\xi\sigma_{13}^2\sigma^{-1}_{11}. 
\end{equation}
To be more intuitive, we rewrite the state preparation conditions $\xi=(1,1/2,0)$ into the gain domain as $g_A=(g_p,g_F,g_x)$, and plot the curves of $g_p$, $g_F$, and $g_x$ versus effective reflectivity $R$ in Fig.~\ref{fig2} (b).

By using the above precise mapping in a concrete example, we illustrate the Wigner functions of the resulting mechanical NGSs caused by a two-photon EPS~\cite{SM}, as shown in Figs.~\ref{fig2}(c)--\ref{fig2}(j). Figures~\ref{fig2}(c)--\ref{fig2}(f) are the resulting mechanical states generated with $R=0.9$ ($\tau\approx3$ ns). Instead of directly performing photon subtraction [Fig.~\ref{fig2}(c)], a prior amplifier allows one to selectively obtain the $P_M$-direction squeezed CS [Fig.~\ref{fig2}(d)], the two-phonon state [Fig.~\ref{fig2}(e)], and the $X_M$-direction squeezed CS [Fig.~\ref{fig2}(f)] with fidelity $\mathcal{F}>0.98$, by choosing the corresponding gains $g_A$ that are shown in the inset of Fig.~\ref{fig2}(b). Furthermore, we show the resulting states generated with $R=0.5$ ($\tau\approx19$ ns) in Figs.~\ref{fig2}(g)--\ref{fig2}(j), where the EPR steering reaches its maximal value. It shows that a longer duration will lead to an increased squeezing and decoherence effect of our desired mechanical NGSs. Here the two more-separated two ``cats'' [Fig.~\ref{fig2}(h)] on $P_M$ and the two closer ``cats'' [Fig.~\ref{fig2}(j)] on $X_M$ are obtained with fidelity $\mathcal{F}> 0.88$. 
% ---------------------------------------------------------------------------
% ---------------------------------------------------------------------------
% ---------------------------------------------------------------------------
\begin{figure}[tbp]
	\centering
	\includegraphics[width=1\linewidth]{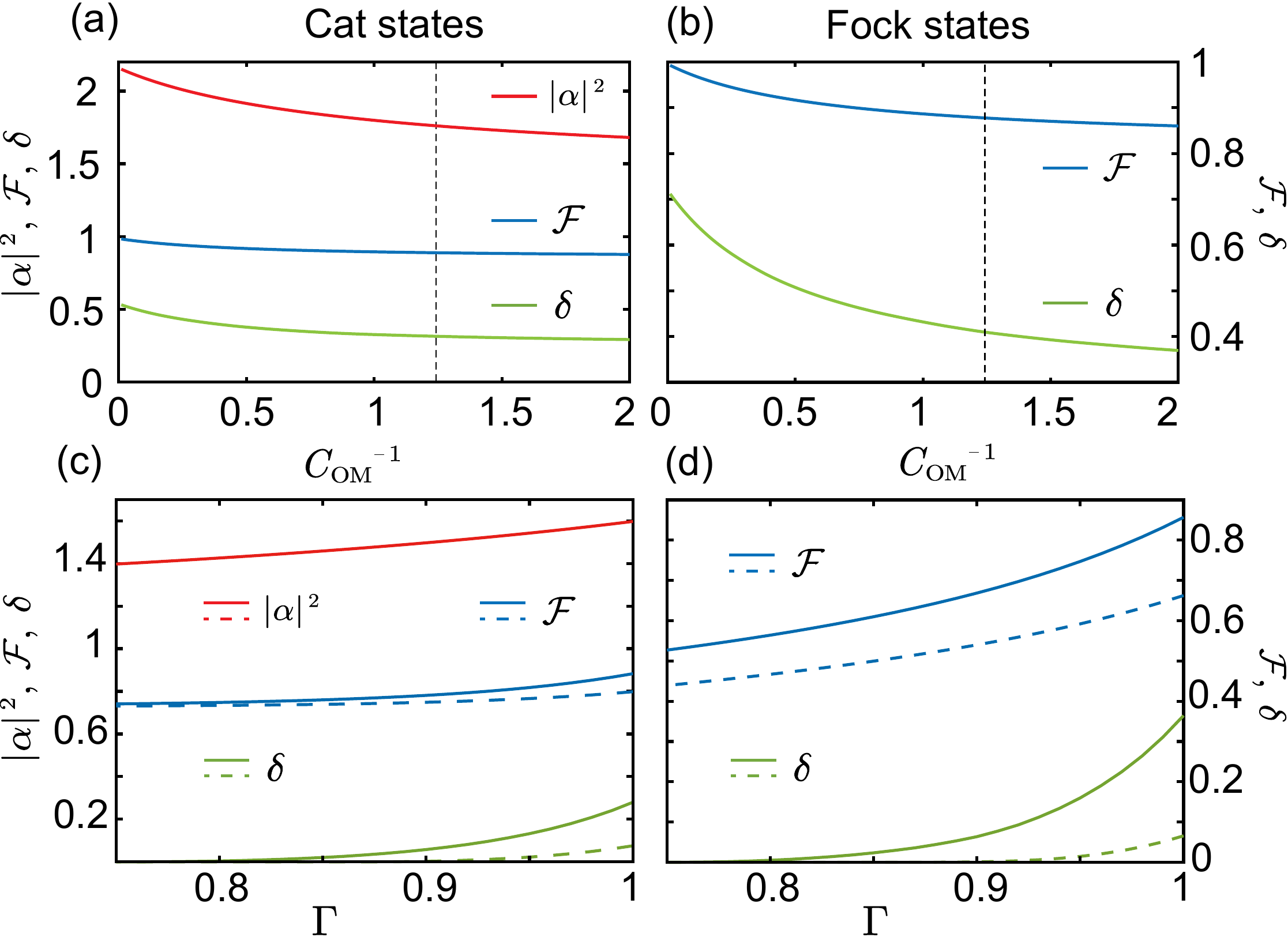}
	\caption{(a), (b) The fidelity $(\mathcal{F})$ and Wigner negativity $(\delta)$ of two resulting NGSs vs $C_{\rm OM}^{-1}=\gamma\kappa/g^2$. The black dashed line indicates $C_{\rm OM}=0.8$. (c), (d) The two resulting NGSs' qualities vs the total optical detection efficiency $\Gamma=\eta+\mu$. Solid (dashed) lines correspond to the amplification noise $n_A=0$ ($n_A=0.1$), and $|\alpha|^2$ is the size of the resulting CSs.}
	\label{fig3}
\end{figure}

\emph{Feasibility and imperfections.} In experiments, the amplifier $\mathcal{U}$ can be implemented by a cavity (with frequency at $\omega_c$) that contains a $\chi^{(2)}$ gain medium~\cite{Lv2015}, as shown in Fig.~\ref{fig1}(a). By pumping the gain medium with driving frequency $\omega_d=2\omega_c$, amplitude $\Lambda$ and phase $\Phi_d$. The Hamiltonian of the cavity is $H_A=i\chi^{(2)}\Lambda(a^2e^{i\Phi_d}-{a^\dagger}^2e^{-i\Phi_d})$ in a frame rotating with $\omega_c$. The phase $\Phi_d=\omega_cl/c$ is used to match the phase of the propagated optical field. From the input-output theory, the amplitude quadrature $X_C$ of the optical field passing through the cavity will obtain a gain, 
\begin{equation}
	g_A=(1+\chi^{(2)}\Lambda/\kappa_A)^2/(1-\chi^{(2)}\Lambda/\kappa_A)^2,
\end{equation}
where $\kappa_A$ is the decay rate of the cavity mode, and the gain $g_A$ can be modulated via the external driving for a desired $\xi$ that determines the mechanical NGSs. However, the noise of practical amplification process and the photon loss during the optical pulse transits to the amplifier both lead to a reduction of the photon-phonon entanglement. Here we quantify these two imperfections by optical transmission efficiency $\eta$ and amplification noise $n_A$ that describes an incoherent squeezing $s=(1+n_A)/g_A$ of phase quadrature $P_C$ when the amplitude quadrature $X_C$ obtains a gain $g_A$.

In addition, the usual photon subtraction event is implemented by a beam splitter with high transmittance which has a low probability of success~\cite{Wenger2004,Wakui2007,Parigi2007,Ra2020}. To address this issue, we introduce the deterministic photon subtraction implemented using the Rydberg-blockade effect~\cite{Honer2011}, and the corresponding experiment of three-photon subtraction has recently been achieved by cascaded cold atomic ensembles~\cite{Stiesdal2021}. The dark counts of the photon subtraction are considered here, which leads to a negligible imperfection. 

Finally, an imperfect homodyne detection with an efficiency $\mu$ is considered. By taking into account all of the above imperfections for the generation of mechanical NGSs with two-photon EPS, we analytically obtain the final Wigner function of the mechanical state,
\begin{equation}
	\begin{aligned}
		&W_{\rho_M}\approx\mathcal{N}\exp(-a X_M^2-b P_M^2)\times\\
		&[(2F_+-2e-2f)^2-4(e-f)F_--\lambda],
	\end{aligned}	
\end{equation}
where $F_{\pm}=(cX_M)^2\pm(dP_M)^2$ and $\lambda=e^2+f^2+6ef$. The Wigner function is described by the six parameters $a-f$ that relate to the total optical detection efficiency $\Gamma=\mu+\eta$ and the amplification noise $n_A$, and their analytical expressions can be found in the Supplemental Material~\cite{SM}.

In order to demonstrate the performance of the protocol with the finite cooperativity and the above imperfections, we numerically evaluate the values of the quality of two resulting NGSs versus the reciprocal of the optomechanical cooperativity $C_{\rm OM}^{-1}=\gamma\kappa/g^2$ in Figs.~\ref{fig3}(a) and \ref{fig3}(b), with a fixed reflectivity $R=0.5$. Additionally, the impacts of the total optical detection efficiency $\Gamma$ and the amplification noise $n_A$ on the quality are illustrated in Figs.~\ref{fig3}(c) and \ref{fig3}(d). The quality of the resulting states is qualified by the following: the fidelity $\mathcal{F}$, the size of mechanical CSs $|\alpha|^2$, and the Wigner negativity $\delta$ that is defined as~\cite{Sun2021}
\begin{equation}
	\delta=\iint_{-\infty}^{\infty}[|W(X_M,P_M)|-W(X_M,P_M)]d X_M d P_M.
\end{equation}
Here a larger value of $\delta$ and $|\alpha|^2$ indicate a better quantum feature and more distant two coherent states, respectively, and $\delta=0$ means that the non-classicality has vanished. 

This result shows that our protocol is surprisingly robust to mechanical decay. Note that since a larger mechanical decay corresponds to a shorter interaction duration $\tau$ for a fixed reflectivity $R$, this can reduce mechanical decoherence but also decrease the size of the CSs. In addition, the quality of the resulting states, not surprisingly, decreases with the increase of various imperfections. However, the quality of the resulting states can be further improved by decreasing the propagation distance of the cavity output light and the input squeezing level. Furthermore, the impact of imperfection in the homodyne is minor here since the homodyne detection efficiency for optical field is usually close to unit, and we find that high-fidelity CSs can be well prepared with an arbitrary outcome, once the direction of the measurement is fixed to $X_C$~\cite{SM}.

\begin{figure}[tbp]
	\centering
	\includegraphics[width=1.0\linewidth]{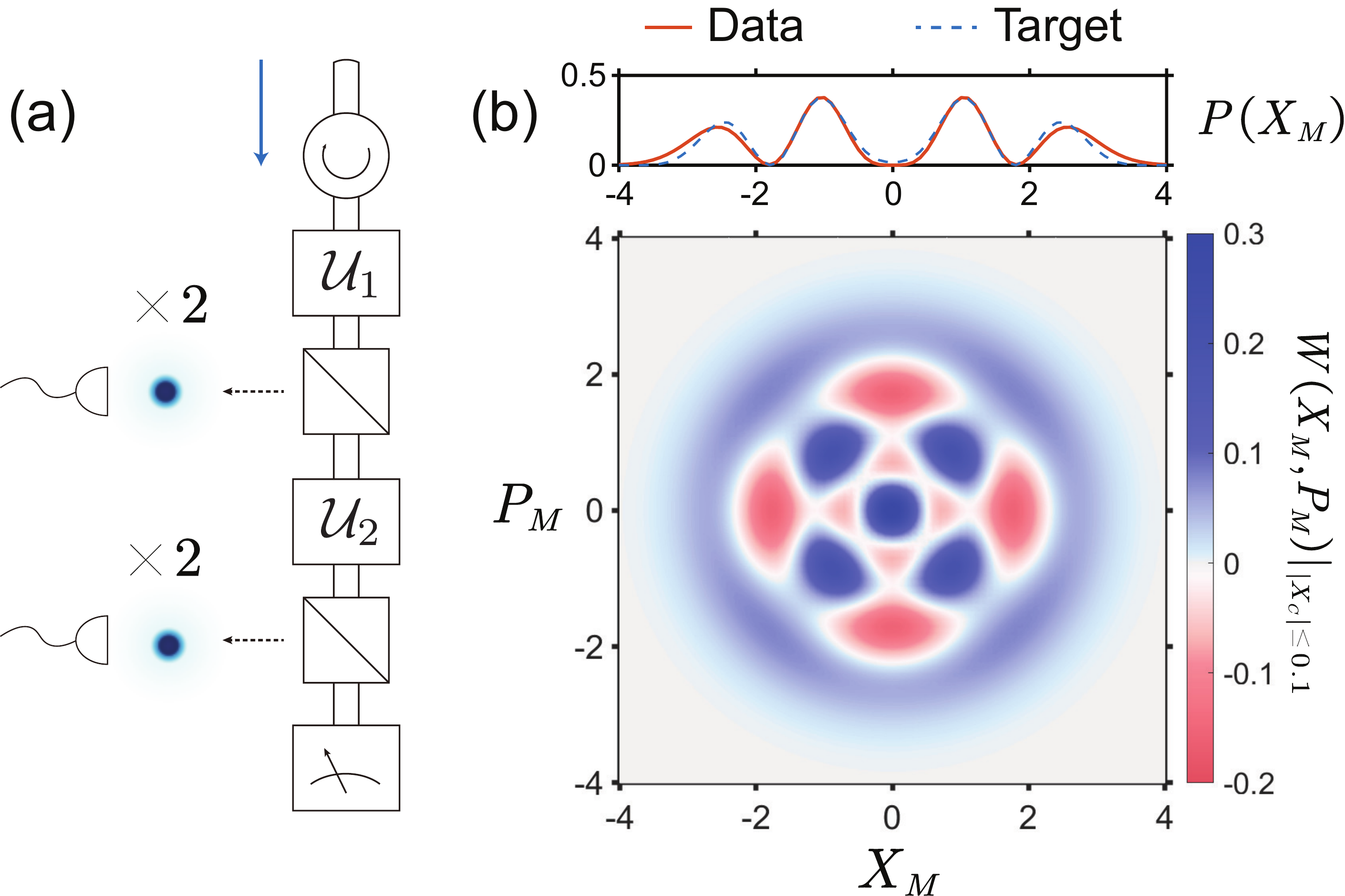}
	\caption{(a) Sketch for preparing a four-component CS at the mechanical mode. (b) The resulting state with a high fidelity $\sim0.98$, where the four cats locate at the positions with $\alpha=1.6e^{in\pi/4}, n=1,3,5,7$. $P(X_M)$ is the probability distribution of the quadrature $X_M$. Here $S_{\rm in}=-6$ dB, $R=0.9$, and $C_{\rm om}=0.8$.}
	\label{fig4}
\end{figure}

% ---------------------------------------------------------------------------
% ---------------------------------------------------------------------------
\emph{Extension and discussion.} This protocol can also be extended to rapidly prepare a four-component CS, which is widely used as cat codes in quantum error correction, by simply performing the two-photon EPS twice on the optical mode $(C)$, as shown in Fig.~\ref{fig4} (a). This construction allows the resulting mechanical NGSs to be manipulated by both $\mathcal{U} (g_{A1})$ and $\mathcal{U}(g_{A2})$. With the conditions $\xi_1=1$ and $\xi_2=-2$, or $\xi_1=0$ and $\xi_2=3$, the mechanical mode will collapse into a four-component state, $\Sigma^4_{n=1}|1.6e^{i(2n-1)\pi/4}\rangle$, after the projective measurement, as shown in Fig.~\ref{fig4}(b) (see Supplemental Material~\cite{SM}). The fidelity of this state reaches a value of $\sim 0.98$, which far exceeds the corresponding fidelity in the achieved experiments~\cite{Ofek2016}. This extension implies that the process of EPS can be repeated and designed for preparing advanced NGSs, which provides an attractive strategy for implementing complex operations on NGSs. 

Since the squeezing technology, photon subtraction, and homodyne measurement employed here are well developed in various quantum systems, the current protocol is also promising to be promoted to other physical platforms, including superconducting circuits~\cite{You2011,Devoret2013,Wendin2017,Blais2021}, spin ensembles~\cite{MA2011,Luca2018}, etc. For example, in superconducting systems working in the microwave domain, field squeezing (the strength can reach $\sim -10$ dB)~\cite{Castellanos2008}, single-photon subtraction and microwave homodyne measurement (with efficiency $\sim0.5$) have been achieved~\cite{Joo2016,Mallet2011}. Note that, in addition to optical systems, superconducting circuits require a low environment temperature to get a negligible thermal occupation.

% ---------------------------------------------------------------------------
% ---------------------------------------------------------------------------
% ---------------------------------------------------------------------------

\emph{Conclusion.} We have shown how mechanical non-Gaussian states can be faithfully generated and manipulated in weak-coupling cavity optomechanical systems $(C_{\rm om}={g^2}/{\kappa\gamma}<1)$. The key ingredient is to use a squeezed pulse with a short duration to rapidly generate the EPR steering and an additional amplifier on the entangled optical mode to remotely control the mechanical states. Our protocol is robust to the mechanical dissipation and promises to deterministically generate mechanical cat states with the photon subtraction implemented by Rydberg atomic ensembles. With the parameters in state-of-the-art experiments, we simulated the preparation of various large-size mechanical non-Gaussian states, all of which maintain a high fidelity. Furthermore, the successful preparation of the four-component cat state convinces us that the cascaded construct of multiple EPSs may be used to generate high-quality advanced NGSs. Such capabilities can be promoted to various quantum systems and may have a wide-ranging impact on future quantum information processing strategies.

% ---------------------------------------------------------------------------
% ---------------------------------------------------------------------------

\emph{Acknowledgments.}---We thank Z. Peng for stimulating discussions. This work is supported by the National Key R\&D Program of China (Grant No. 2019YFA0308200) and the National Natural Science Foundation of China (Grant No. 11874432). Q.H. thanks the National Natural Science Foundation of China (Grants No. 11975026 and No. 12125402), Beijing Natural Science Foundation (Grant No. Z190005) and the Key R\&D Program of Guangdong Province (Grant No. 2018B030329001) for financial support. 
F.-X. S. acknowledges the National Natural Science Foundation of China (Grants No. 12147148) and the China Postdoctoral Science Foundation (Grant No. 2020M680186). J.-Q. L. is supported in part by National Natural Science Foundation of China (Grants No. 12175061, No. 12247105, and No. 11935006) and the Science and Technology Innovation Program of Hunan Province (Grants No. 2021RC4029 and No. 2020RC4047).
% ---------------------------------------------------------------------------
% ---------------------------------------------------------------------------
%%%%%%%
\bibliography{reference_FGMNGS.bib}
\end{document}

\title{\large Supplementary material for: \\Fast Generation of High-Fidelity Mechanical Non-Gaussian States via Additional Amplifier and Photon Subtraction}
\author{Dong-Long Hu}
\affiliation{School of Physics, Sun Yat-sen University, Guangzhou 510275, China }
\author{Jia-Jin Zou}
\affiliation{School of Physics, Sun Yat-sen University, Guangzhou 510275, China }
\author{Feng-Xiao Sun}
\affiliation{State Key Laboratory for Mesoscopic Physics, School of Physics, Frontiers Science Center for Nano-Optoelectronics, and Collaborative Innovation Center of Quantum Matter, Peking University, Beijing 100871, China}
\author{Jie-Qiao Liao}
\affiliation{Key Laboratory of Low-Dimensional Quantum Structures and Quantum Control of Ministry of Education,
	Key Laboratory for Matter Microstructure and Function of Hunan Province, Department of Physics and
	Synergetic Innovation Center for Quantum Effects and Applications, Hunan Normal University, Changsha 410081, China}
\author{Qiongyi He}
\affiliation{State Key Laboratory for Mesoscopic Physics, School of Physics, Frontiers Science Center for Nano-Optoelectronics, and Collaborative Innovation Center of Quantum Matter, Peking University, Beijing 100871, China}
\author{Ze-Liang Xiang}
\affiliation{School of Physics, Sun Yat-sen University, Guangzhou 510275, China }
\date{\today}

\maketitle
% ---------------------------------------------------------------------------
% ---------------------------------------------------------------------------

\section{Dynamics of red-detuned optomechanical coupling}

We consider the mechanical mode of a resonator coupled to the optical cavity mode, connected with the waveguide, through the radiation pressure. This coupling can be described by the Hamiltonian ($\hbar=1$)
\begin{equation}
	H=\omega_{c}c^\dagger c+\omega_m m^\dagger m+g_0c^\dagger c(m+m^\dagger).
\end{equation}
where $g_0$ is the single-photon optomechanical coupling strength, which is generally very weak. The cavity mode is driven by a coherent field, with frequency $\omega_p$. In a frame rotating with $\omega_p$, we obtain
\begin{equation}
	H_1=\Delta c^\dagger c+\omega_m m^\dagger m+g_0c^\dagger c(m+m^\dagger)+i\sqrt{2\kappa}\varepsilon_p(c-c^\dagger).
\end{equation}
Here $\Delta=\omega_{c}-\omega_p$ is the frequency detuning, and $\varepsilon_p=\sqrt{P/\hbar\omega_p}$ is the strength of the driving laser. The driving field makes the cavity mode can be written as the sum of a mean-field and the quantum fluctuation $c=\beta+\delta c$. The dynamics of the cavity mode and the mechanical mode can then be described by quantum Langevin equations (QLEs) as
\begin{align}
	&\dot{\beta}=(-i\Delta-\kappa)\beta-\sqrt{2\kappa}\varepsilon_p,\\
	\label{S4}
	&\delta\dot{c}=(-i\Delta-\kappa)\delta c-ig_0\beta(m+m^\dagger)-\sqrt{2\kappa}\delta c_{\rm in}(t),\\
	\label{S5}
	&\dot{m}=(-i\omega_m-\gamma)m-ig_0(\beta\delta c^\dagger+\beta^*\delta c)-\sqrt{2\gamma}m_{in}(t).
\end{align}
Here $\kappa$ ($\gamma$) is the rate that the cavity (mechanical) mode decaying to the waveguide (environment), and $c_{\rm in}(t)$ [$m_{\rm in}(t)$] is a $\delta$-correlated noise operator from the waveguide (environment), which satisfies $[c_{\rm in}(t),c^\dag_{\rm in}(t')]=\delta(t-t')$ \{$[m_{\rm in}(t),m^\dag_{\rm in}(t')]=\delta(t-t')$\}. The mean field is $\beta\approx{-i\sqrt{2\kappa}\varepsilon_p}/{\Delta}$ ($\kappa\ll\Delta$). We consider the red-detuned field ($\Delta=\omega_m$), which leads to the parametric term $\delta cm$ can be ignored under the rotating-wave-approximation. Therefore, we can obtain an effective Hamiltonian as ($\delta c\rightarrow c$)
\begin{equation}
	H_{\rm eff}=\Delta c^\dagger c+\omega_m m^\dagger m-ig(c^\dagger m-cm^\dagger).
\end{equation}
where $g=g_0|\beta|$ is the linearized optomechanical coupling strength, with $|\beta|={\sqrt{2\kappa}\varepsilon_p}/{\Delta}$ being the light amplitude of the cavity mode. This Hamiltonian gives rise to a beam-splitter-type interaction between the optical mode and the mechanical mode. We now discuss the dynamics of this system when a probe field is pulsed into the waveguide with a finite duration $\tau$. In a framework rotating with $\Delta$, QLEs~(\ref{S4}) and (\ref{S5}) reduce to 
\begin{align}
	&\dot{c}(t)=-\kappa c(t)-g m(t)-\sqrt{2\kappa} c_{\rm in}(t),\\
	\label{s8}
	&\dot{m}(t)=-\gamma m(t)+g c(t)-\sqrt{2\gamma}m_{\rm in}(t).
\end{align}
The cavity decay rate is assumed to $\kappa\gg g$, which allows one to make the bad-cavity approximation $c\approx-\frac{g}{\kappa}m-\sqrt{\frac{2}{\kappa}}c_{in}(t)$. Then, the effective dynamics of the system can be obtained from Eq.~(\ref{s8}) and the input-output relation $c_{\rm out}=c_{\rm in}+\sqrt{2\kappa}c$, as 
\begin{align}
	&c_{\rm out}(t)=-\sqrt{\frac{2g^2}{\kappa}} m(t)-c_{\rm in}(t),\\
	&\dot{m}(t)=-G m(t)-\sqrt{\frac{2g^2}{\kappa}}c_{\rm in}(t)-\sqrt{2\gamma}m_{\rm in}(t),
\end{align}
where $G={g^2}/{\kappa}+\gamma$ is the effective decay rate of the mechanical mode. It can be directly understood as the sum of two decay channels, the waveguide and the environment. For an input pulse with a finite duration. It is convenient to use a set of normalized temporal modes to describe these two modes, namely
\begin{align}
	&C_{\rm in}=\sqrt{\frac{2GR}{T}}\int^{\tau}_0c_{\rm in}(s)e^{Gs}ds, \ \ \ \ M_{\rm in}=m(0),\\
	&C_{\rm out}=\sqrt{\frac{2G}{T}}\int^{\tau}_0c_{\rm out}(s)e^{-Gs}ds,\ \ \  \ M_{\rm out}=m(\tau),
\end{align}
where $R=e^{-2G\tau}$ is the effective reflectivity of the beam-splitter, and $T=1-R$ is the transmissivity. At the end of the pulse, the beam-splitter-type scattering relation reads as
\begin{align}
	\label{S13}
	&M_{\rm out}=M_{\rm in}\sqrt{R}-\sqrt{\frac{T}{2G}}\left(\sqrt{\frac{2g^2}{\kappa}}C_{\rm in}+\sqrt{2\gamma}M_{m}\right),\\
	\label{S14}
	&C_{\rm out}=-\left[\sqrt{2\gamma}\sqrt{\frac{T}{2G}}M_{\rm in}+\tilde{C}_{\rm in}+\frac{g^2}{ \kappa G}(C_{\rm in}\sqrt{R}-\tilde{C}_{\rm in})-\sqrt{{\frac{g^2\gamma}{\kappa G^2}}}(\tilde{M}_{m}-\sqrt{R}M_m)\right],
\end{align}
with a set of the assist normalized temporal modes, namely
\begin{align}
	\tilde{C}_{\rm in}&=\sqrt{\frac{2G}{T}}\int_{0}^{\tau}c_{\rm in}(s)e^{-Gs}ds,\nonumber\\
	M_m&=\sqrt{\frac{2GR}{T}}\int_{0}^{\tau}m_{\rm in}(s)e^{Gs}ds,\nonumber\\
	\tilde{M}_{m}&=\sqrt{\frac{2G}{T}}\int_{0}^{\tau}m_{\rm in}(s)e^{-Gs}ds.
\end{align}
By defining quadratures $X_o={(o+o^\dagger)}/{\sqrt{2}}$ and $P_o={(o-o^\dagger)}/{i\sqrt{2}}$, the scattering relation~(\ref{S13})~(\ref{S14}) can be rewritten as 
\begin{align}
	&X_{M,\rm out}=X_{M,\rm in}\sqrt{R}+\sqrt{\frac{T}{2G}}\left(-\sqrt{\frac{2g^2}{\kappa}}X_{C,\rm in}-\sqrt{2\gamma}X_{M,m}\right),\nonumber\\
	&P_{M,\rm out}=P_{M,\rm in}\sqrt{R}+\sqrt{\frac{T}{2G}}\left(-\sqrt{\frac{2g^2}{\kappa}}P_{C,\rm in}-\sqrt{2\gamma}P_{M,m}\right),\\
	&X_{C,\rm out}=-\left[\sqrt{\frac{g^2T}{\kappa G}}X_{M,\rm in}+X_{\tilde{C},\rm in}+\frac{g^2}{\kappa G}(X_{C,\rm in}\sqrt{R}-X_{\tilde{C},\rm in})+{{\sqrt{\frac{g^2\gamma}{\kappa G^2}}}}(\sqrt{R}X_{M,m}-X_{\tilde{M},m})\right],\nonumber\\
	&P_{C,\rm out}=-\left[\sqrt{\frac{g^2T}{\kappa G}}P_{M,\rm in}+P_{\tilde{C},\rm in}+\frac{g^2}{\kappa G}(P_{C,\rm in}\sqrt{R}-P_{\tilde{C},\rm in})+{{\sqrt{\frac{g^2\gamma}{\kappa G^2}}}}(\sqrt{R}P_{M,m}-P_{\tilde{M},m})\right].
\end{align}
\section{Squeezing induced photon-phonon entanglement}

The quantum state $\rho_{\rm out}$ of the system at $t=\tau$ depends on the statistics of the operators at the initial moment. $m_{\rm in}(t)$ is the thermal noise arising from the surrounding environment which can be described by correlation function  $\langle m_{\rm in}(t)m^\dagger_{\rm in}(t')\rangle=(1+n_m)\delta(t-t')$, where $n_m$ is the thermal occupation of the mechanical mode. For creating the expected photon-phonon entanglement to remotely manipulate the mechanical mode, we input a squeezed vacuum into the waveguide, and $c_{\rm in}(t)$ is the squeezing noise that can be described by $\langle X_{c_{\rm in(t)}}X_{c_{\rm in(t')}}\rangle=S_{\rm in}/2\delta(t-t')$ and  $\langle P_{c_{\rm in(t)}}P_{c_{\rm in(t')}}\rangle=S_{\rm in}^{-1}/2\delta(t-t')$, where $S_{\rm in}$ is the squeezing strength of the input pulse and we assume that the squeezing of the input pulse has a temporal uniformity, hence $\langle X_{\rm C_{in}(\tau)}X_{\rm C_{in}(\tau)}\rangle=\langle X_{\rm c_{in}(0)}X_{\rm c_{in}(0)}\rangle=S_{\rm in}/2$. We consider the mechanical mode is initially prepared at the ground state, and the system is at a low temperature $T$ with a negligible thermal occupation $n_{m}=1/[\exp(\hbar\omega_{m}/k_BT)-1]$. By using these correlation functions, we can obtain the analytical expression of the covariance matrix V
\begin{equation}
	\label{S18}
	V=\begin{pmatrix}
		V_M&V_{MC}\\
		V_{MC}^T&V_C
	\end{pmatrix},
\end{equation}
with the matrix elements
\begin{align}
	\label{S19}
	&V_M=\frac{1}{2}\left\{\frac{T g^2}{\kappa G}\textbf{S}+\left[R+\frac{\gamma T}{G}\right]\textbf{I}(1+2n_m)\right\},\\
	\label{S20}
	&V_{MC}=\sqrt{\frac{g^2 TR}{4\kappa G}}\left\{\left(\frac{g^2}{\kappa G}+\frac{\gamma}{G}\frac{2G\tau}{T}\right)\textbf{S}-\left[1+\frac{\gamma}{G}\left(\frac{2G\tau}{T}-1\right)\right]\textbf{I}(1+2n_m)\right\},\\
	\label{S21}
	&V_{C}=\frac{1}{2}\left\{\left(\frac{g^4R}{\kappa ^2G^2}+\frac{\gamma^2}{G^2}+\frac{g^2\gamma}{\kappa G^2}\frac{4G\tau R}{T}\right)\textbf{S}+\left[\frac{g^2 T}{\kappa G}+\frac{g^2\gamma}{\kappa G^2}\left(R+1-\frac{4G\tau R}{T}\right)\right]\textbf{I}(1+2n_m)\right\},
\end{align}
where $\textbf{S}/2=\{S_{\rm in},0;0,S^{-1}_{\rm in}\}/2$ is the initial covariance matrix of the optical mode, $\textbf{I}$ is the identity matrix and $(1+2n_m)\textbf{I}/2$ is the initial covariance matrix of the mechanical mode, while $V_{\tau=0}=1/2((1+2n_m)\textbf{I}\oplus\textbf{S})$. By using the covariance matrix $V=\{V_M,V_{MC};V_{MC}^T,V_C\}$, the photon-phonon entanglement can now be quantified in terms of the logarithmic negativity $E_N=\max\{0,-\ln2\nu\}$, where $\nu=\sqrt{\mu(V)-[\mu(V)^2-4\det V]^{1/2}}/\sqrt{2}$ and $\mu(V)=\det V_M+\det V_C-2\det V_{MC}$~\cite{Adesso2004}. Moreover, the EPR steering from the mechanical mode to the cavity output mode is given by $\mathcal{G}^{M\rightarrow C}=\max\{0,\frac{1}{2}\ln(\det V_M/\det V)\}$~\cite{Kogias2015}. In the main text, we illustrate the curves of the $E_N$ and $\mathcal{G}^{M\rightarrow C}$ corresponding to various reflectively $R$ with the situation ($n_m=0,C_{\rm om}=0.8$). The value of the entanglement depends on the input squeezing strength and the reflectivity of the effective beam-splitter, and the peaks always appear around at $R=\frac{1}{2}$ because the mixedness between mechanical mode and optical mode here is maximal. Note that the squeezing-induced entanglement is applicable for general beam-splitter-type coupling in various physical platforms.

\begin{figure}[tbp]
	\centering
	\includegraphics[width=0.8\linewidth]{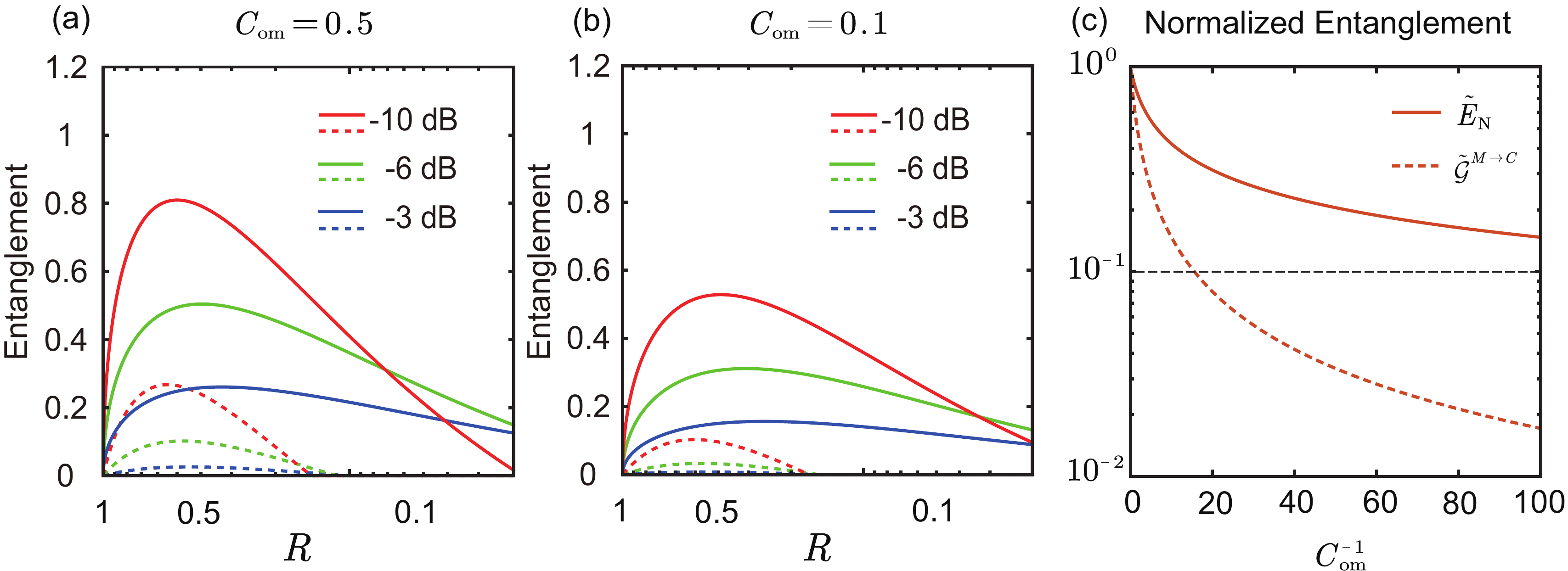}
	\caption{The logarithmic negativity $E_N$ and EPR steering with low optomechanical cooperativity, (a) $C_{\rm om}=0.5$ and (b) $C_{\rm om}=0.1$. The normalized correlations are various with $C_{\rm om}$ (c). The correlations are produced with $R=0.5$ and $S_{\rm in}=-6$ dB and are normalized by the  corresponding value with $C_{\rm om}^{-1}=0$.}
	\label{fig.s1}
\end{figure}

We now discuss the squeezing-induced entanglement with the low optomechanical cooperativity $C_{\rm om}$. It only affects the ratio of the coupling between mechanical mode and the pulse of $g^2/G\kappa=C_{\rm om}/(C_{\rm om}+1)$. Thus the low value of $C_{\rm om}$ only leads to a decreasing value of the produced photon-phonon correlations but does not hinder their generation. In the FIG.~\ref{fig.s1}. (a) and (b), we illustrate the correlations produced with $C_{\rm om}<1$, where the correlations still appear whenever $S_{\rm in}\neq0$ even for $C_{\rm om}=0.1$ and the value of these correlations will increase with the squeezing strength of the input pulse. In addition, the influence of $C_{\rm om}$ on the correlations' values is shown in FIG.~\ref{fig.s1} (c). Here the values decrease along the optomechanical cooperativity, while the normalized EPR steering attenuates to $0.1$ when $C_{\rm om}=0.055$. Especially, the decreasing normalized logarithmic negativity $>0.1$ even for $C_{\rm om}=0.01$. It supports that the generation of photon-phonon entanglement using our method is robust to low optomechanical cooperativity, which provides the precondition for remotely producing mechanical NGSs, and even the optomechanical system in the weak-coupling regime. 

\section{Effect of Engineered photon subtraction}

In the main text, we have shown that EPS performed on entangled optical mode $C_{\rm out}$ can be utilized to control the type and the size of the resulting mechanical NGSs $\rho_M$. Here we show the derivations of the precise mapping between the process of the EPS and $\rho_M$. Note that the next analysis and conclusions are applicable for arbitrary non-displaced bipartite Gaussian entangled states. We start from the two-mode Gaussian entanglement state $\rho_{\rm out}$. Thus, we can conveniently obtain the Wigner function of $\rho_{\rm out}$,
\begin{equation}
	W_G(V)=\frac{1}{2\pi\sqrt{\det V}}\exp(-1/2\textbf{u}V^{-1}\textbf{u}^T).
\end{equation}
Here $\textbf{u}=(X_M,P_M,X_C,P_C)$. Firstly, the phase-sensitive amplification of $X_C$ can be described as a symplectic matrix $U=\{I,0;0,\mathcal{E}\}$ in the phase space, where $\mathcal{E}=\{g_A,0;0,1/g_A\}$ and $g_A$ is the gain of the amplifier $\mathcal{U}$. Thus, the amplifier $\mathcal{U}$ transfers the covariance matrix $V$ into $U V U^T$. Then, the deterministic multi-photon subtraction can be realized by going, one by one, through the identical system with the Rydberg atomic ensembles~\cite{Stiesdal2021}. This process transfers the two modes Gaussian state $\mathcal{U}\rho_{\rm out}\mathcal{U}^\dag$ into an entangled non-Gaussian state $\rho^{-n}\propto C^n\mathcal{U}\rho_{\rm out}\mathcal{U}^\dag (C^\dag)^n$, where $C$ is the annihilation operator of the optical mode $C_{\rm out}$. In the phase space, the single-photon subtraction can be described as a second-order differential operator $\hat{\mathcal{C}}$,
\begin{equation}
	W_{C\rho C^\dag}\Leftrightarrow \hat{\mathcal{C}}W_{\rho}=\frac{1}{2}(X_C^2+P_C^2+1+X_C\partial_{X_C}+P_C\partial_{P_C}+\frac{1}{4}\partial^2_{X_C}+\frac{1}{4}\partial^2_{P_C})W_{\rho}.
\end{equation}
Lastly, the projective measurement can be described as a projective operator $\hat{\Pi}=|X_C=0\rangle\langle X_C=0|$, and thus one can obtain the resulting mechanical NGS $\rho_M$ and the corresponding Wigner function of $\rho_M$,
\begin{align}
	&\rho_M=\langle\textbf{E}(g_A,n)\rho_{\rm out}\textbf{E}^\dag(g_A,n)\rangle_{X_C=0},\\
	\label{S25}
	&W_{\rho_{M}}=\lim_{\epsilon\rightarrow0}\iint dX_CdP_C\hat{\mathcal{C}}^nW_G(UVU^T)\frac{e^{-(X_C)^2/2\epsilon^2}}{\sqrt{2\epsilon^2}},
\end{align}
where $\textbf{E}(g_A,n)=C^n\mathcal{U}(g_A)$ is a combined operator that describes the EPS, and $\varepsilon$ is the measurement error of the projective measurement. Eq.~(\ref{S25}) is still valid when $\varepsilon\ll 1$, we show the results with $\varepsilon=0.1$ in the main text. In addition, we also provide the analytical expression of the Wigner function of the non-Gaussian entangled state after the atom ensemble absorbed n photons from the optical mode $C_{\rm out}$, 
\begin{equation}
	W_G^{-n}(V)=\hat{\mathcal{C}^n}W_G(V)\propto Q_n W_G(V),
\end{equation}
where $Q_n(X_M,P_M,X_C,P_C)$ is a 2n-th order polynomial that can be obtained by
\begin{align}
	&Q_0=1,Q_1=\left\{1-\frac{\sigma_{33}+\sigma_{44}}{2}+L_X^2+L_P^2\right\},\\
	&Q_n=Q_1Q_{n-1}-L_X\frac{\partial Q_{n-1}}{\partial X_C}-L_P\frac{\partial Q_{n-1}}{\partial P_C}+\frac{1}{4}\left(\frac{\partial^2 Q_{n-1}}{\partial^2 X_C}+\frac{\partial^2 Q_{n-1}}{\partial^2 P_C}\right),(n\geq2),
\end{align}
with $L_X=(\sigma_{33}-1)X_C+\sigma_{13}X_M$ and  $L_P=(\sigma_{44}-1)P_C+\sigma_{24}P_M$, and $\sigma=1/2V^{-1}$. Then one can obtain the analytical expression of the Wigner function of the resulting mechanical NGS $\rho_M$ from Eq.~(\ref{S25}). So far, we show how to obtain the analytical expression of $\rho_{\rm M}$ from the covariance matrix $V$ in phase space.
\begin{figure}[tbp]
	\centering
	\includegraphics[width=0.8\linewidth]{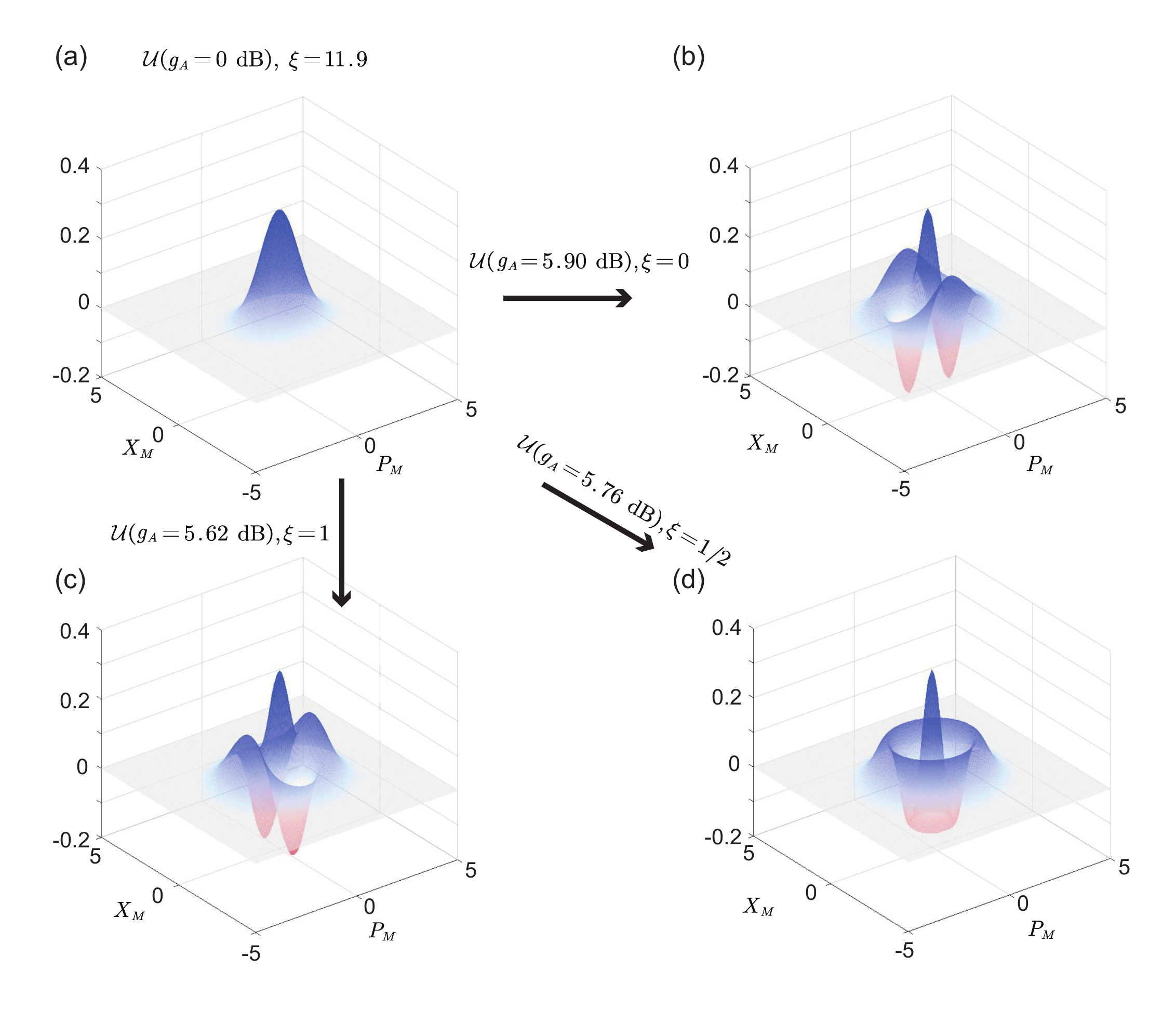}
	\caption{The effect of the amplifier $\mathcal{U}$. (a) A resulting mechanical state is generated by directly performing 2-photon subtraction and the projective measurement. A resulting mechanical state is generated with $\xi=0$ (b), $\xi=1/2$ (c), $\xi=1$ (d). Here $S_{\rm in}=-6 \rm dB$, $R=0.9$ and $C_{\rm om}=0.8$.}
	\label{fig.S2} 
\end{figure}
Especially if the intrinsic decay rate of mechanical mode is vanished ($\gamma=0$), the Wigner function and the wave function of the $\rho_{\rm out}$ have the relation,
\begin{align}
	&|\psi(X_M,X_C)|^2=\iint d P_M d P_C W_G( V ),\\
	&|\tilde{\psi}(P_M,P_C)|^2=\iint d X_M d X_C W_G( V ),
\end{align}
which establishes a connection between the covariance matrix $V$ and the wave function of the entangled Gaussian state $\rho_{\rm out}$,
\begin{align}
	&\psi(X_M,X_C)=e^{i\phi_1} \exp[-\frac{1}{2}(\sigma_{11}X_M^2+\sigma_{33}X_C^2+2\sigma_{13}X_M X_C)],\\
	&\tilde{\psi}(P_M,P_C)=e^{i\phi_2} \exp[-\frac{1}{2}(\sigma_{22}P_M^2+\sigma_{44}P_C^2+2\sigma_{24}P_M P_C)],
\end{align}
where $\phi_1$ and $\phi_2$ are global phases that is irrelevant. By performing the EPS and the following projective measurement, the relation between the $X$-wave functions of $\rho_{\rm out}$ and $\rho_M$ reads as
\begin{equation}
	\psi(X_M)\propto(X_C+\partial_{X_C})^n\psi(X_M,X_C/\eta)|_{X_C=0},
\end{equation}
where $(X_C+\partial_{X_C})/\sqrt{2}$ is the $X$ representation of the annihilation operator $C$, with $C=(X_C+i P_C)/\sqrt{2}$ and $P_C=-i\partial_{X_C}$. In the main text Eqs.~(3) and (4), we have given the analytical expression of the $\psi(X_M)$,
\begin{align}
	&\psi(X_M)\propto\phi_{n,\xi}(X_M)\exp\left(-\frac{X_M^2}{2\sigma_{11}^{-1}}\right),\\
	\label{S35}
	&\phi_{n,\xi}(X_M)\propto\sigma_{13}^n\Sigma_k^{[n/2]}\frac{(-1)^kn!2^{n-2k}}{k!(n-2k)!}\left(\frac{X_M}{\sqrt{2\sigma_{11}^{-1}}}\right)^{n-2k}{\xi}^{k}. (\sigma_{13}\neq0)
\end{align}
Here $\xi=(\sigma_{33}-g_A)/\sigma_{13}^2\sigma_{11}^{-1}$ can be remotely modulated by the amplifying strength $g_A$ of the operation $\mathcal{U}$, and the subtracted photon number $n$ actually decides the size of the mechanical NGSs. Eq.~(\ref{S35}) also can be rewritten as 
\begin{equation}
	\phi_{n,\xi}(X_M)\propto\xi^{n/2}\sigma_{13}^n H_{n}\left(\frac{X_M}{\sqrt{2\xi\sigma_{11}^{-1}}}\right), (\xi\neq0,\sigma_{13}\neq0),
\end{equation}
where $H_n(x)$ is the n-th Hermite polynomial. We define the performance of the single-photon subtraction in our protocol that is the remote photon-phonon conversion rate. $\Gamma={\rm Tr}(\rho_Mm^\dag m)/n|_{\sigma_{11}=1}$ can be obtianed as
\begin{equation}
	\Gamma=\int d X_M \psi^\dag(X_M)(X_M-\partial_{X_M})(X_M+\partial_{X_M})\psi(X_M)/2n|_{\sigma_{11}=1},
	\label{S37}
\end{equation}
where $\sigma_{11}=1$ indicates the elimination of the contribution of the mechanical squeezing effect to phonon number. $\Gamma$ is an axisymmetric function whose axis (maximal value) is $\xi=1/2$ ($\Gamma=1$). We define good performance of the photon subtraction as $\Gamma>0.5$, i.e., less than 50\% performance is failed. Therefore, one can choose an appropriate $g_A$ to make $\xi\sim1/2$ for enhancing the performance of the photon subtraction. Furthermore, We find
\begin{equation}
	\psi(X_M)\propto\left\{
	\begin{aligned}
		&X_M^n \exp\left(-\frac{X_M^2}{2\sigma^{-1}_{11}}\right)\Leftrightarrow\tilde{\psi}(P_M)\propto H_n\left(\frac{P_M}{\sqrt{2\sigma_{11}}}\right)\exp\left(-\frac{P_M^2}{2\sigma_{11}}\right),~~ \xi=0,\\
		&H_n\left(\frac{X_M}{\sqrt{\sigma_{11}^{-1}}}\right)\exp\left(-\frac{X_M^2}{2\sigma^{-1}_{11}}\right)\Leftrightarrow\tilde{\psi}(P_M)\propto H_n\left(\frac{P_M}{\sqrt{\sigma_{11}}}\right)\exp\left(-\frac{P_M^2}{2\sigma_{11}}\right),~~ \xi=1/2,\\
		&H_n\left(\frac{X_M}{\sqrt{2\sigma_{11}^{-1}}}\right)\exp\left(-\frac{X_M^2}{2\sigma^{-1}_{11}}\right)\Leftrightarrow\tilde{\psi}(P_M)\propto P_M^n \exp\left(-\frac{P_M^2}{2\sigma_{11}}\right),~~ \xi=1.
	\end{aligned}
	\right.
\end{equation}
%
%  ---------------------------------------------------------------------------
%
\begin{figure}[b]
	\centering
	\includegraphics[width=0.85\linewidth]{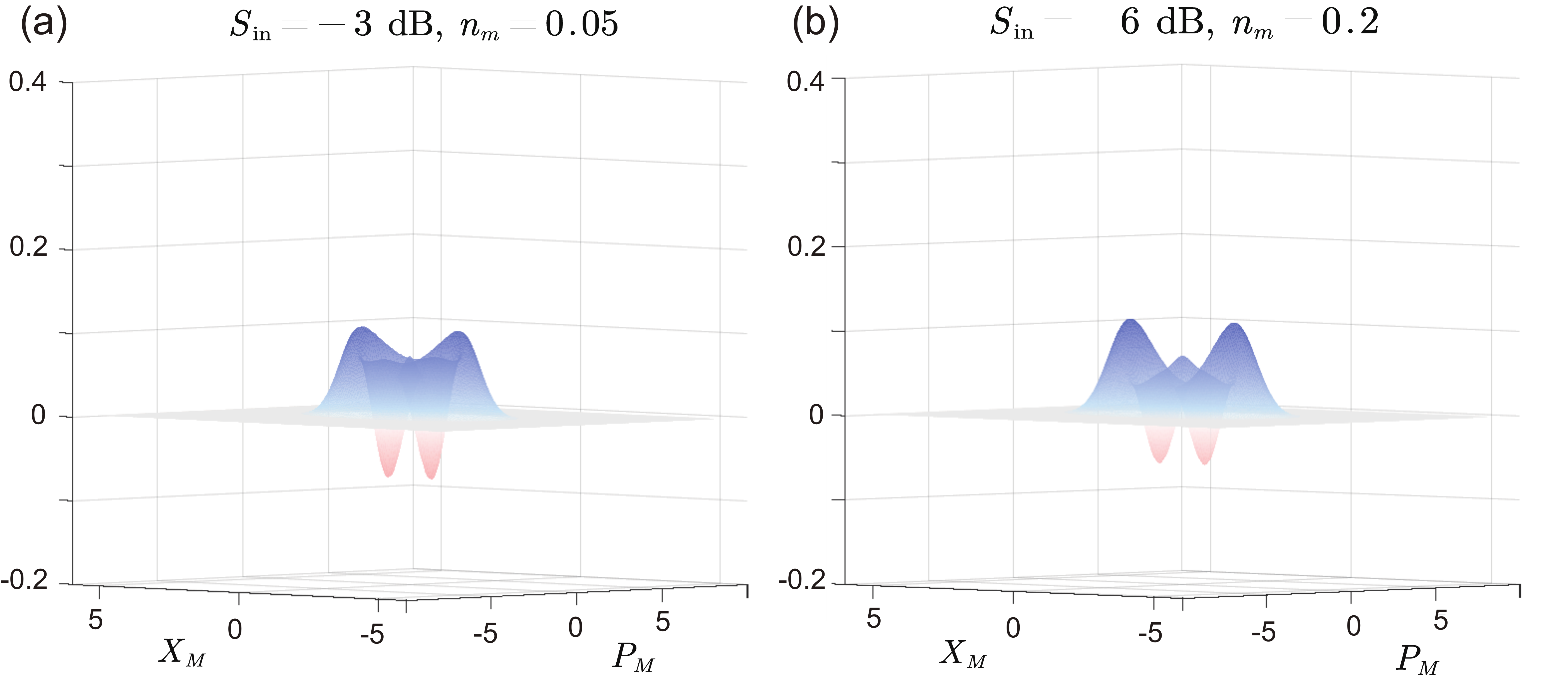}
	\caption{The resulting states with the thermal occupation of the mechanical state. (a) [(b)] the squeezing strength $S_{\rm in}=-3$ [-6] dB, the thermal occupation $n_m=0.05~[0.2]$ and the amplifying strength $g_A=2.88~[5.66]$ dB.  Here $R=0.9$ and $C_{\rm om}=0.8$.}
	\label{fig.s3} 
\end{figure}

As the Ref~\cite{Ourjoumtsev2007}, the wave function in the form of $X_M^n \exp(-{X_M^2}/{2\sigma_{11}})$ corresponds to a squeezed cat state with a high fidelity $\mathcal{F}\approx1-0.03/n$, since the wave function has two peaks located at the $X_M=\pm\sqrt{n\sigma_{11}^{-1}}$, which corresponds to the superposition of two coherent states with a coherent amplitude $\alpha=\sqrt{n}$ and a squeezing strength $(2\sigma_{11})^{-1}$. The parity of the wave function is as same as $n$, and thus it corresponds to an odd (even) cat state when $n$ is odd (even). The squeezing effect of the mechanical mode will increase with the pulse duration $\tau$ by $\sigma_{11}=R+TS_{\rm in}^{-1}$. Though utilizing the precise mapping between the EPS and the resulting mechanical NGS $\rho_M$, one can employ the appropriate gain $g_A$ of the amplifier $\mathcal{U}$ to prepare the desired type of the mechanical NGS, as shown in FIG.~\ref{fig.S2}. Here, for example, one can choose $\xi=0$ (satisfied by $g_A= 5.90~\rm dB$) and $\xi=1$ (satisfied by $g_A= 5.62~\rm dB$) to make the two possible ``cats" located at $X_M$ and $P_M$ direction, respectively. Besides, choosing $\xi=1/2$ (satisfied by $g_A=5.76~ \rm dB$) creates a mechanical Fock state. However, the absence of the amplifier $\mathcal{U}$, i.e., directly performing photon subtraction only obtains a terrible result. It supports that the amplifier $\mathcal{U}$ extremely enhances the performance of the remote preparation of mechanical NGSs. Note that the preparation of mechanical NGSs using squeezing-induced entanglement has no strict requirements for the squeezing strength $S_{\rm in}$ (also the amplifying strength $g_A$ ). If the mechanical mode is initially in the ground state, the larger $S_{\rm in}$ only gives a larger mechanical squeezing effect at the same time ($\sigma_{11}=R+TS_{\rm in}^{-1}$), which we will show in the following. 

At the last in this section, we discuss the effect of the thermal occupation of the mechanical mode. In the recent experiment\cite{Mirhosseini2020}, the resonance frequency of the mechanical mode has reached $\omega_{m}/2\pi\sim 5$ GHz. The resonator is embedded in the dilution refrigerator with temperature $\sim 15$ mK~\cite{Mirhosseini2020} and thus a negligible thermal occupation. The situation with $n_m\approx0$ has been discussed in the main text. Here we consider a thermal occupation $n_m=0.05$ (95\% in the ground state) with the temperature $T=0.1$ K.  Eqs.~(\ref{S18}--\ref{S21}) provide the covariance matrix $V$ with the thermal occupation, which we utilize to illustrate the Wigner function of $\rho_M$ produced by 2-photon EPS and the successful projective measurement, with a low input squeezing strength $S_{\rm in}=-3$ dB, as shown in the FIG.~\ref{fig.s3} (a). Here the desired mechanical CS can still be effectively generated with a fidelity $\mathcal{F}=0.62$, the size $|\alpha|^2=2.1$, and the $\delta=0.20$, where the required amplifying strength $g_A=2.88$ dB. In addition, the larger squeezing of the input pulse can give a higher tolerance of the thermal occupation, as shown in FIG.~\ref{fig.s3} (b). Here we employ $S_{\rm in}=-6$ dB that allows the cat state can be generated with $n_m=0.2$ (83\% in the ground state), where the fidelity $\mathcal{F}=$ 0.70, the size $|\alpha|^2=2.1$, and the Wigner-negativity $\delta=0.18$ are reached by $=5.66$ dB. The reason for the phenomenon is that the thermal occupation will interfere the generation of the photon-phonon entanglement, and a larger squeezing of the input allows the quantum steering to produce with a larger thermal occupation.
\begin{figure}[b]
	\centering
	\includegraphics[width=0.5\linewidth]{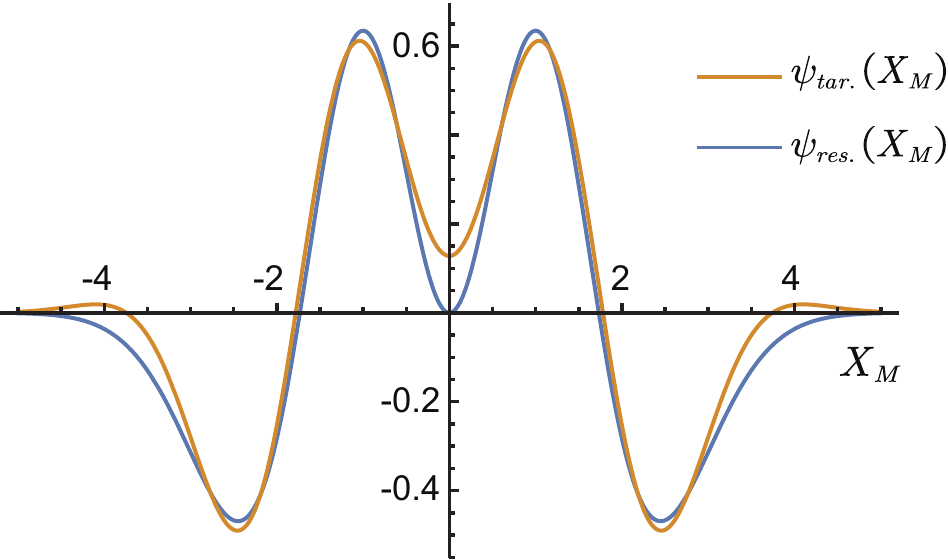}
	\caption{The wave function of the ideal four-component cat state (orange line) and the resulting wave function of the mechanical mode, which is generated by $S_{\rm in}=-6$ dB and $R=0.9$ (blue line). The overlap between such two states is $\sim 0.98$.}
	\label{fig.s4}
\end{figure}
\section{Generation of Four component cat states}
In this section, we show how to prepare the four-component cat state by two cascaded 2-photon EPSs. Eq.~(\ref{S35}) tells us that a phase-sensitive amplifier before the multi-photon subtraction can adjust the shape of the resulting states with a parameter $\xi$. It is natural to think that performing more than one squeezing operation on the optical output mode may obtain diversified quantum states with more controllable parameters. Based on this idea, we try to obtain the wave function of $\rho_M$ by performing two cascaded 2-photon EPS [$\textbf{E}(g_{A2},2)\textbf{E}(g_{A1},2)$] on the optical output mode. With $\gamma=0$, the relation between the resulting mechanical states $\rho_M$ and the $\rho_{\rm out}$ is $\rho_M=\langle\textbf{E}(g_{A2},2)\textbf{E}(g_{A1},2)\rho_{\rm out}[\textbf{E}(g_{A2},2)\textbf{E}(g_{A1},2)]^\dag\rangle_{X_C=0}$. The resulting mechanical state can be directly obtained as
\begin{align}
	&\psi(X_M)\propto\sigma_{13}^4\left[\left(\frac{X_M}{s}\right)^4-(5\xi_1+\xi_2)\left(\frac{X_M}{s}\right)^2+2\xi_1^2+\xi_1\xi_2\right]e^{-\frac{X_M^2}{2s^2}},\\
	&\tilde{\psi}(P_M)\propto\sigma_{13}^4[({s P_M})^4-(5\xi_1+\xi_2-6)({s P_M})^2+2\xi_1^2+\xi_1\xi_2-(5\xi_1+\xi_2-6)+3]e^{-\frac{P_M^2s^2}{2}},
\end{align}
with the controllable parameters $\xi_1=({\sigma_{33}-g_{A1}})/{\sigma_{13}^2\sigma_{11}^{-1}}$ , $\xi_2=({\sigma_{33}-g_{A2}g_{A1}})/{\sigma_{13}^2\sigma_{11}^{-1}}$ and $s=\sqrt{\sigma_{11}^{-1}}$. Due to the four-component cat state has $C_4$ symmetry in the phase space, the wave function $\psi(X_M)$ and $\psi(P_M)$ should have a similar form, while the squeezing strength $s^2=\sigma_{11}^{-1}=(R+TS_{\rm in}^{-1})^{-1}$ is negligible with a short duration (i.e., $R=0.9$). Thus we have the condition $5\xi_1+\xi_2=3$. Especially, by choosing the previously mentioned values for $\xi_1=0,1/2,1$, we can obtain the corresponding $\xi_2=3,1/2,-2$, respectively. The wave functions of these combinations are
\begin{equation}
	\psi(X_M)\propto\left\{
	\begin{aligned}
		&(X_M^4/s^4-3X_M^2/s^2) \exp\left(-\frac{X_M^2}{2s^2}\right)\Leftrightarrow\tilde{\psi}(P_M)\propto (P_M^4s^4-3P_M^2s^2) \exp\left(-\frac{P_M^2s^2}{2}\right),~~ \xi_1=0,~~\xi_1=1,\\
		&H_4\left(\frac{X_M}{s}\right)\exp\left(-\frac{X_M^2}{2s^2}\right)\Leftrightarrow\tilde{\psi}(P_M)\propto H_4({P_M}s)\exp\left(-\frac{P_M^2s^2}{2}\right),~~ \xi_1=1/2,\\
	\end{aligned}
	\right.
\end{equation}
where we find the combination $\xi_1=0,\xi_2=3$ and $\xi_1=1,\xi_2=-2$ correspond to a same wave function that has a high fidelity $\sim 0.98$ with respect to the ideal four-component cat state $\Sigma^4_{n=1}|1.6e^{i(2n-1)\pi/4}\rangle$, as shown in FIG.~\ref{fig.s4}. It can be understood as the effect of the second EPS divides each cat of the cat states produced by the first 2-photon EPS into two new cats. In addition, the combination $\xi_1=\xi_2=1/2$ corresponds to a four-phonon state, where the gain of the second amplifier $\mathcal{U}_2$ is 0 dB, i.e., degenerates to the one 4-photon EPS situation. By using the parameters, in a realistic optomechanical system, the preparation process of the four-component cat state can be described as 
\begin{equation}
	W_{\rho_M}=\iint_{\varepsilon=0.1} d X_C d P_C \hat{\mathcal{C}}^2 \mathcal{U}(g_{A2}) \hat{\mathcal{C}}^2 W[U(g_{A2}) VU_1^T(g_{A1})] (\hat{\mathcal{C}}^\dag)^2\mathcal{U}^\dag(g_{A2})(\hat{\mathcal{C}}^\dag)^2 \exp(-x^2/2\varepsilon^2)/\sqrt{2\varepsilon^2}.
\end{equation}
This relation is utilized to numerically calculate the Wigner function of the resulting mechanical four-component cat state in the main text, as shown in Fig.~(4). This state preparation process implies that the different combinations of the EPSs can be used to produce more complicated NGSs. Therefore, we can predict that a 2n-component cat state can potentially be generated by cascading n 2-photon EPSs in our protocol.

\section{feasibility and imperfections}

\begin{figure}[tbp]
	\centering
	\includegraphics[width=1\linewidth]{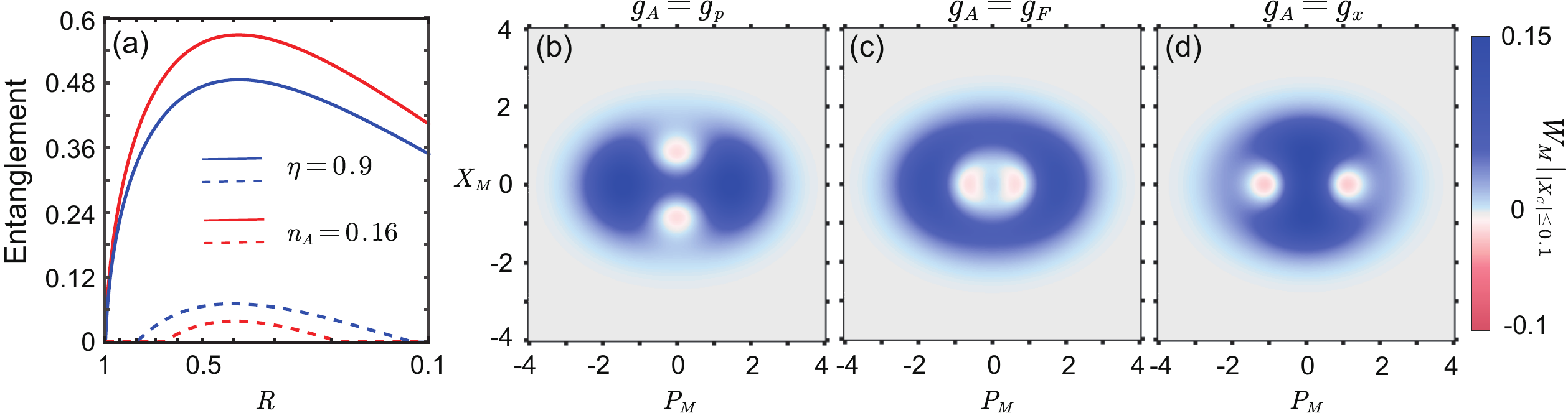}
	\caption{(a) The logarithmic negativity (solid line) and the EPR steering (dashed line) versus effective reflectivity $R$ with transmission efficiency $\eta=0.9$ (blue line) and amplification noise $n_A=0.16$ (red line), respectively. (b)-(d) The Wigner functions of the  mechanical states produced with the transmission loss, while the amplification noise leads to similar results. Here the dark counts $\nu=0.98$ of photon subtraction and homodyne efficiency $\mu=0.8$ have been considered. The other parameters are $C_{\rm om}=0.8$ and $R=0.5$. }
	\label{fig.s5}
\end{figure}

First of all, we consider the practical loss during the cavity output field transits to a distant site for conditional NGSs preparation. This process can be described by mixing the entangled optical mode and vacuum noise on a beam-splitter with transmission efficiency $\eta$. This results a transformation of the matrix $\sigma$, as
\begin{align}
	&\sigma_{11}\mapsto\sigma_{11}-(1-\eta)\sigma_{13}^2/(\eta+(1-\eta)\sigma_{33}),\\
	&\sigma_{22}\mapsto\sigma_{22}-(1-\eta)\sigma_{24}^2/(\eta+(1-\eta)\sigma_{44}),\\
	&\sigma_{33}\mapsto\sigma_{33}/(\eta+(1-\eta)\sigma_{33}),\\
	&\sigma_{44}\mapsto\sigma_{44}/(\eta+(1-\eta)\sigma_{44}),\\
	&\sigma_{13}\mapsto\sqrt{\eta}\sigma_{13}/(\eta+(1-\eta)\sigma_{33}),\\
	&\sigma_{24}\mapsto\sqrt{\eta}\sigma_{24}/(\eta+(1-\eta)\sigma_{44}).
\end{align} 
Such a process introduces the decay of the photon-phonon entanglement. For a short distance, we take a transmission efficiency as a reasonable value $\eta=0.9$ in numerical simulation of entanglement, as shown in FIG.~\ref{fig.s5}(a).

Further, we discuss the amplification process. The physical implementation of the OPA is that a cavity (with resonance frequency $\omega_c$) contains a $\chi^{(2)}$ gain medium that is pumped with driving frequency $\omega_d=2\omega_c$, amplitude $\Lambda$ and phase $\Phi_d$. The Hamiltonian of the cavity is $H_A=i\chi^{(2)}\Lambda(a^2e^{i\Phi_d}-{a^\dagger}^2e^{-i\Phi_d})$ in a frame rotating with $\omega_c$. The phase $\Phi_d=\omega_cl/c$ is used to offset the phase shift caused by the optical field propagation. From the input-output theory, the amplitude quadrature $X_C$ of the transit field will be amplified with a gain 
\begin{equation}
	g_A=(1+\chi^{(2)}\Lambda/\kappa_e)^2/(1-\chi^{(2)}\Lambda/\kappa_e)^2.
\end{equation}
Here $\kappa_e$ is the OPA cavity decay rate, and $g_A$ can be controlled by adjusting the amplitude $\Lambda$ of the pumping field. The amplification of optical field with noise gives rise to $V\mapsto UVU^T$ with the matrix $U=\{I,0;0,\mathcal{E}\}$ where $\mathcal{E}=\{g_A,0;0,(1+n_A)/g_A\}$. The noise usually is negligible when the gain $g_A$ is not large. Here we show the entanglement generated with $n_A=0.16$ in Fig.~\ref{fig.s5}(a), namely, if the amplitude quadrature $X_C$ obtains a gain $g_A=x$ dB, the phase quadrature $P_C$ will obtain a inverse gain $-x+0.6$ dB. It shows that the transmission loss and amplification noise lead to similar effects for the entanglement decay, i.e., reducing the entanglement values and also forbid the EPR steering to be generated by a very short input.

Then we adopt cascaded Rydberg atom ensembles to implement the multi-photon subtraction events. Strong van der Waals interactions between Rydberg atoms in an ensemble lead to a blockade effect that only supports a single Rydberg collective excitation. In addition, the enhanced collective photon-atom coupling $\Omega_N=\sqrt{N}g_0\Omega_C/2\Delta_C$ allows an ensemble to efficiently absorb one photon from the incoming light field. Then the excitation will be rapidly dephasing into a manifold of $N-1$ collective dark states to suppress the photon re-emission of the ensembles. Such a special physical mechanism is promising to achieve a deterministic photon subtraction. Here $N$ is the number of atoms in an ensemble, $g_0\Omega_C/2\Delta_C$ is single-photon-atom coupling rate~\cite{Honer2011}. In experiments, a click of the ion detection of the ensembles heralds a successful photon subtraction event, which immediately transfers the quantum state of the entangled system into $c\rho c^\dagger$, and a typical dark counts $\nu=0.98$ for the detection have been considered in calculation.

Subsequently, a homodyne detection measures the amplitude quadrature $X_C$ of the optical mode with an efficiency $\mu$, which projects the quantum state into $\rho_M|_{X_C=\zeta}$ with a outcome $\zeta$. In the main text, we concentrate the good projection $|X_C|\leq\epsilon$ with a measurement error $\epsilon=0.1$. By considering these imperfections the Wigner-function of the final mechanical non-Gaussian states with 2-photon EPS can be obtained as
\begin{align}
	W_M=&\mathcal{N}\exp(-a X_M^2-b P_M^2)[(2F_+-2e-2f)^2-4(e-f)F_--\lambda].
\end{align}
Here $F_{\pm}=(cX_M)^2\pm(dP_M)^2$ and $\lambda=e^2+f^2+6ef$, the Wigner function can be described by the six parameters 
\begin{align}
	&a=\sigma_{11} + ((1 - \mu + \epsilon^2)\sigma_{13}^2)/(\mu(-1 + \sigma_{33}) - (1 + \epsilon^2)\sigma_{33}),\\
	&b=\sigma_{22}-\sigma_{24}^2/\sigma_{44},\\
	&c=(1 + \epsilon^2)\sigma_{13}/(\mu + \sigma_{33} - \mu\sigma_{33} + \epsilon^2\sigma_{33}),\\
	&d=\sigma_{22}/\sigma_{44},\\
	&e=(1 + \epsilon^2)(-1 + \sigma_{33}))/(\mu + \sigma_{33} - \mu\sigma_{33} + \epsilon^2\sigma_{33}),\\
	&f=(\sigma_{44}-1)/\sigma_{44}.
\end{align}

We illustrate the Wigner-functions of resulting mechanical states in FIGs.~\ref{fig.s5}(b)-(d) to show the impacts of the transmission loss on our results, while the impacts of amplification noise is similar. It shows that the $P_M$-direction CS with qualities $|\alpha|^2=1.7$, $\delta=0.04$, $\mathcal{F}=0.76$ [FIG.~\ref{fig.s5}(b)], the squeezed Fock states with qualities $\delta=0.05$, $\mathcal{F}=0.66$, $s=-2$ dB [FIG.~\ref{fig.s5}(c)], and the $X_M$-direction squeezed CS with qualities $|\alpha|^2=1.7$, $\delta=0.06$, $\mathcal{F}=0.81$, $s=-4.6$ dB [FIG.~\ref{fig.s5}(d)] are achieved with transmission efficiency $\eta=0.9$. Here we consider a homodyne efficiency $\mu=0.8$, however, the homodyne efficiency for optical field usually close to unit, which can further improve the quality of NGSs. These results demonstrate that our protocol is promising to prepare and manipulate mechanical NGSs with Wigner-negativity in practical weak-coupling optomechanical systems. 

\begin{figure}[tbp]
	\centering
	\includegraphics[width=0.8\linewidth]{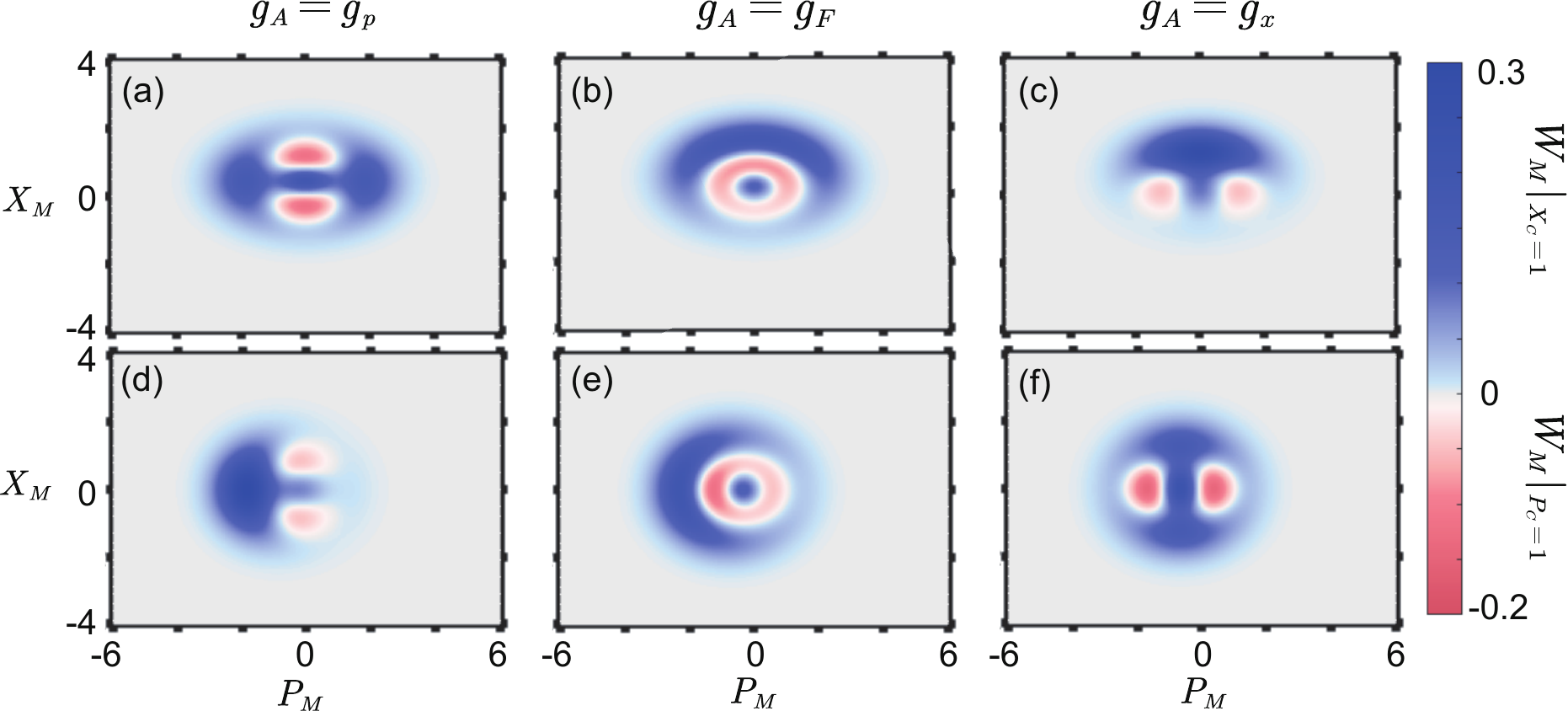}
	\caption{(a)-(c) The resulting mechanical states with the outcome of the projective measurement $X_C=1$. (d)-(f) The resulting mechanical states with the outcome of the projective measurement $P_C=1$. Here $C_{\rm OM}=0.8$ and $R=0.5$.}
	\label{fig.s6}
\end{figure}

Lastly, we show the effect of the imperfect projective measurement with outcome $X_{\theta}=\zeta$. Above mechanical NGSs are obtained with the outcome $|X_{C}|\leq0.1$ of the projective measurement, which usually has a limited probability and the analogs has been discussed in previous works\cite{Ourjoumtsev2007,Sun2021}. However, we find that such a constraint is no longer necessary for our protocol. With the fixed projective measurement direction $X_{\theta}=X_{C}$ $(X_{\theta}\equiv X_C\cos\theta+P_C\sin\theta)$, which is easily realized in experiments by fixing the phase between the local oscillator and the signal light, the arbitrary measuring outcome $X_C=\zeta$ gives to a mechanical CS when $g_A=g_p$, as 
\begin{align}
	&\psi(X_M)\propto H_n[(X_M-d)/\sqrt{2\sigma_{11}^{-1}}]\exp\left(-\frac{(X_M-d)^2}{2\sigma_{11}^{-1}}\right),\\
	&\tilde{\psi}(P_M)\propto \exp(idP_M)P_M^n \exp\left(-\frac{P_M^2}{2\sigma_{11}}\right).
\end{align}
The wave functions correspond to a mechanical CS with two coherent states on $P_M$, while the different outcome $\zeta$ only gives to a displacement $d=-\zeta\sigma_{13}/(\sqrt{g_A}\sigma_{11})$ of $X_M$ in the phase space. In the upper row of FIG.~\ref{fig.s6}, we illustrate the resulting states with the outcome $X_C=1$ and the 2-photon EPS, where the desired $P_M$-direction mechanical CS still emerges with a fidelity $\sim0.9$, as shown in FIG.~\ref{fig.s6} (a). Similarly, the $X_M$-direction mechanical CS with high fidelity is also still present with the outcome $P_C=1$ of the projective measurement, as shown in FIG.~\ref{fig.s6} (f). This indicates the generation of high-fidelity mechanical CSs with $\xi=1$ or $\xi=0$ is not affected by the projective results if one fixed the measuring direction, which leads to a deterministic generation of the mechanical cat states when the subtracted photons are detected. Excitingly, the efficiency of the process can be extremely improved by using the deterministic photon subtraction induced by the Rydberg-blockade effect. Additionally, the imperfect projective measurement will decrease the quality of the mechanical squeezed Fock states, as shown in FIGs.~\ref{fig.s6} (b) and (e). Here the fidelity of these squeezed Fock states still can reach $\mathcal{F}>0.7$. 
\bibliography{reference_FGMNGS.bib}